# Multi-Objective Aerodynamic Optimization of Ride Height and Rake Angle in a Sedan Car Using CFD and Machine Learning

**Mahdi Kheirkhah[1], Ehsan Roohi[*2], Mahmoud Pasandidehfard[1]**

[1] Mechanical and Industrial Engineering, University of Massachusetts Amherst, 160 Governors Dr., Amherst, 01003, Massachusetts, USA
[2] Department of Aerospace Engineering, University of Massachusetts, Amherst, USA
[*]Email: roohie@umass.edu

**Abstract:** This study investigates the aerodynamic performance of an Audi A4 sedan using CFD analysis. A 3D model was developed in SolidWorks and validated against DrivAer Notchback wind-tunnel data, showing only a 3.25% deviation in drag coefficient ($C_d$). Ride height varied from 1.336 to 1.536 m and rake angle from 0° to 5°, across four Reynolds numbers (Re = $4.87×10^6$, $9.75×10^6$, $14.61×10^6$, $19.48×10^6$). Gradient Boosting emerged as the most accurate predictive model ($R^2 \approx 0.97$ for $C_d$ and 0.96 for lift coefficient, $C_l$), outperforming Random Forest and LightGBM. Differential Evolution optimization was performed under balanced, drag-focused, and downforce-focused conditions. Reynolds number had minimal impact on optimum location; therefore, detailed results are reported for Re = $9.75×10^6$, with other Re showing similar trends. The baseline geometry (ride height = 1.436 m, rake angle = 0°) exhibited $C_d$ = 0.313 and $C_l$ = +0.0288. Balanced optimization achieved $C_d$ = 0.287 (−8.31%) and $C_l$ = −0.0826 (≈387% increase in downforce). Minimum drag condition reached $C_d$ = 0.285 (−8.95%) with slight positive lift ($C_l$ = +0.0142), while maximum downforce optimization reached $C_l$ = −0.1084 (≈476% increase) with a 6.71% drag penalty ($C_d$ = 0.334). Near-optimal solutions were found within ride heights of 1.341–1.365 m and rakes of 0.158–4.610°, indicating robust aerodynamic performance. Machine learning predictions were further validated against CFD with <3% error in $C_d$. Finally, aerodynamic optimization could either reduce fuel consumption by up to 9.1% or enhance downforce by over 500%, depending on the selected condition.

## 1. Introduction

In vehicle aerodynamic design, two critical parameters—ride height and rake angle—play a pivotal role in reducing drag force and controlling lift. Adjustments in these parameters can optimize aerodynamic performance across various speeds and flow conditions. In recent years, numerical modeling using software like ANSYS Fluent, combined with machine learning algorithms for data analysis, has become an effective tool for analyzing and optimizing these characteristics [1, 2]. Multiple studies have demonstrated that lowering the vehicle's ride height, especially near the front axle, tends to reduce the drag coefficient. In contrast, excessive rake angle can increase lift and reduce stability [3]. This effect is more pronounced in racing cars, which often exhibit significant positive rake angles [4]. Among standard models used in aerodynamic numerical studies, the DrivAer model, with its three body styles (Notchback, Fastback, Estateback), has been extensively analyzed. Compared to simpler models like the Ahmed body, DrivAer better represents real-world industrial shapes and has gained wide usage recently [5]. For instance, Semeraro and Schito examined the combined effects of ride height and tire deformation on aerodynamic performance, highlighting that neglecting tire deformation when lowering ride height can lead to inaccurate drag estimation [6]. Similarly, Džijan et al. found that in race cars, increasing rake angle and lowering ride height decrease drag but increase lift [4].

Recent advances include the application of intelligent algorithms such as Random Forest, XGBoost, and evolutionary optimization techniques like Differential Evolution for parameter optimization. These methods can learn flow patterns from numerical simulation data to determine the optimal ride height and rake angle for any given Reynolds number, which can be used as a form of active control [7, 8]. Validation of numerical models is crucial in aerodynamic research. Studies that utilize wind tunnel experimental data or published numerical results to validate their models gain higher scientific credibility [9]. For DrivAer-based studies, comparison with experimental data from institutions such as TU Munich forms the basis of model validation. Regarding standard ranges of ground clearance and rake angle, passenger sedans typically have a ground clearance between 120 and 160 mm and a rake angle between 0° and 2°, whereas race cars feature a front ground clearance as low as 50–80 mm and rake angles ranging from 3° to 6° [10, 11]. SUVs or off-road vehicles usually have a ground clearance above 200 mm to ensure obstacle clearance, albeit at the cost of increased drag [12]. Overall, given the complex interplay of vehicle geometry, flow velocity, and ground conditions, the use of validated numerical models, intelligent optimization algorithms, and experimental data is indispensable for effective aerodynamic design optimization.

While prior studies have examined parameters such as ride height and rake angle in isolation, the main innovation of this work is a unified, machine-learning-based framework for the simultaneous, multi-objective optimization of these two key variables. This paper offers several novel contributions: First, it develops an accurate predictive model using a XGBoost Regressor, capable of forecasting drag and lift coefficients for any combination of ride height, rake angle, and Reynolds number with low computational cost. Second, it implements a robust optimization process using Differential Evolution under three weighting strategies: Balanced, Drag-Focused, and Downforce-Focused. We report quantitative identification and analysis of the trade-off between fuel economy (lower drag) and stability (higher downforce). Finally, we demonstrate the important finding that the optimal geometric configuration shows very weak dependence on Reynolds number (vehicle speed) over the investigated range, which has practical implications for the design of active suspension systems.

The remainder of the paper is organized as follows. Section 2 details the numerical methodology, model geometry, and boundary conditions. Section 3 presents and analyzes the simulation and optimization results. Section 4 concludes with key findings and suggestions for future work.

## 2. Governing Equations

### 2.1 Reynolds-Averaged Navier-Stokes (RANS) Framework

The turbulent characteristics of the flow are captured using the Reynolds-Averaged Navier-Stokes (RANS) formulation. Each flow variable, such as the velocity component $u_i$ is split into a mean part and a fluctuating part:

$$u_i = u_i' + \bar{u}_i \tag{1}$$

where $\bar{u}_i$ denotes the time-averaged velocity and $u_i'$ represents the fluctuating component. A similar decomposition can be applied to pressure and other scalar quantities. Substituting these into the continuity and momentum equations results in the following time-averaged forms:

$$\frac{\partial \rho}{\partial t} + \frac{\partial (\rho u_i)}{\partial x_i} = 0 \tag{2}$$

$$\frac{\partial (\rho u_i)}{\partial t} + \frac{\partial \left(\rho u_i u_j\right)}{\partial x_j} = -\frac{\partial \bar{p}}{\partial x_i} + \frac{\partial}{\partial x_j}\left[\mu\left(\frac{\partial u_i}{\partial x_j} + \frac{\partial u_j}{\partial x_i}\right) - \frac{2}{3}\frac{\partial u_k}{\partial x_k}\delta_{ij}\right] + \frac{\partial}{\partial x_j}\left(-\rho \overline{u_i' u_j'}\right) \tag{3}$$

Here, $\rho$ is the fluid density, $\overline{p}$ is the mean pressure, and $\delta_{ij}$ is the Kronecker delta. The last term, involving the Reynolds stress tensor $\overline{u_i'u_j'}$, accounts for turbulence effects. To complete the system, a suitable closure model is required [13].

## 2.2 Boussinesq Hypothesis and Reynolds Stress Approximation

A widely used approach to model the Reynolds stresses is the Boussinesq hypothesis, which approximates the stress components based on the gradients of the mean velocity:

$$-\rho\overline{u_i'u_j'} = \mu_t\left(\frac{\partial u_i}{\partial x_j} + \frac{\partial u_j}{\partial x_i}\right) - \frac{2}{3}\left(\rho k + \mu_t\frac{\partial u_i}{\partial x_i}\right)\delta_{ij} \tag{4}$$

In this expression, $\mu_t$ is the eddy viscosity, and $k$ represents the turbulent kinetic energy. This method forms the basis of several turbulence models, including Spalart-Allmaras, $k$–$\varepsilon$ and $k$–$\omega$ models [13, 14].

## 2.3 The SST $k$–$\omega$ Turbulence Model

Among two-equation models, the $k$–$\omega$ framework is known for its effectiveness in capturing near-wall effects. The transport equation for turbulent kinetic energy is given by:

$$\frac{\partial(\rho k)}{\partial t} + \frac{\partial(\rho u_j k)}{\partial x_j} = P_k - \beta^*\rho k\omega + \frac{\partial}{\partial x_j}\left[(\mu + \sigma_k\mu_t)\frac{\partial k}{\partial x_j}\right] \tag{5}$$

where the production term is defined as:

$$P_k = -\rho\overline{u_i'u_j'}\frac{\partial\overline{u}_i}{\partial x_j} \tag{6}$$

The SST $k$–$\omega$ turbulence model, introduced by Menter [15], combines the strengths of the $k$–$\omega$ formulation near walls with the $k$–$\varepsilon$ approach in the outer boundary layer. This hybrid formulation improves predictions in regions of adverse pressure gradients and flow separation, while reducing sensitivity to freestream turbulence [16]. Comparative studies over canonical geometries, such as the Ahmed body, have demonstrated that the SST $k$–$\omega$ model provides reliable predictions of wake structures, pressure distributions, and aerodynamic coefficients among RANS approaches [17]. Several investigations on the DrivAer model and other realistic vehicle geometries have also adopted SST $k$–$\omega$, showing satisfactory agreement with experimental data [5, 9, 18]. Overall, the SST $k$–$\omega$ model represents a widely accepted and effective choice for automotive aerodynamic simulations. The governing equations for the specific dissipation rate $\omega$ and the turbulent viscosity are as follows:

$$\frac{\partial(\rho\omega)}{\partial t} + \frac{\partial(\rho u_j \omega)}{\partial x_j} = \gamma \frac{\omega}{k} P_k - \beta\rho\omega^2 + (1-F_1)2\rho\sigma_{\omega 2}\frac{1}{\omega}\frac{\partial k}{\partial x_j}\frac{\partial\omega}{\partial x_j} + \frac{\partial}{\partial x_j}\left[(\mu + \sigma_\omega\mu_t)\frac{\partial\omega}{\partial x_j}\right] \quad (7)$$

$$\mu_t = \rho\frac{k}{\max\left(\omega, \frac{F_2 S}{a_1}\right)} \quad (8)$$

The SST $k$–$\omega$ turbulence model incorporates two blending functions, $F_1$ and $F_2$, to achieve a smooth transition between the near-wall and free-stream formulations. The function $F_1$ activates the standard $k$–$\omega$ model near solid boundaries, where it is known to provide accurate resolution of the boundary layer, while gradually switching to the transformed $k$–$\varepsilon$ model away from the wall. The second function, $F_2$, is employed in the eddy-viscosity formulation and ensures that the limiter for turbulent viscosity is only applied in regions where it is needed, thereby preventing excessive turbulent viscosity in free-shear flows. This dual-blending strategy enhances model stability and predictive capability across a wide range of flow conditions and Reynolds numbers. However, it may lead to numerical stiffness in zones with sharp gradients [14, 15].

## 3. Numerical Investigation

## 3.1 Computational Model Geometry

The geometry under investigation corresponds to a passenger vehicle from the Audi automotive group, specifically modeled after the Audi A4. In terms of rear configuration, it matches the standard DrivAer Notchback model [5]. The 3D model of the Audi A4 was created using the blueprint method within the SolidWorks 2020 software environment. As illustrated in Figure 1(a), the body of the Audi vehicle is displayed from multiple perspectives. The DrivAer Notchback model, shown in Figure 1(b), is a widely used generic vehicle geometry developed by the Institute of Aerodynamics and Fluid Mechanics at the Technical University of Munich, intended to support aerodynamic investigations of passenger cars [18, 19]. Among the various rear-end configurations, the notchback is characterized by a distinct cut-off shape at the rear, which promotes earlier flow separation and consequently leads to a greater pressure drop in the wake region compared to fastback designs. This increased separation results in a higher drag coefficient [20, 21]. However, the same flow behavior can also be beneficial for generating aerodynamic downforce. After passing over the rear window, the airflow must change direction and impinge upon the rear deck, which causes a rise in static pressure in that region. Compared to the smoother flow over fastback geometries, this pressure recovery effect in notchback designs enhances rear-end stability and controllability at higher speeds [22].

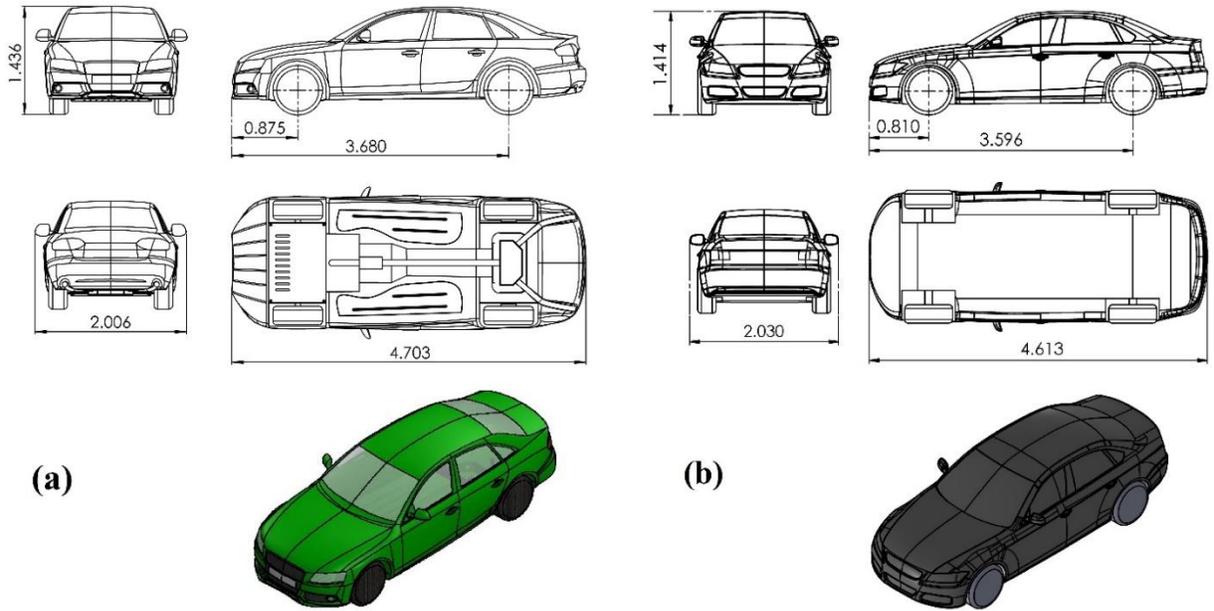

**Figure 1.** The two-dimensional schematic representations include: (a) the designed body of the Audi A4, modeled in SolidWorks 2020 based on blueprint drawings (dimensions in meters); and (b) the DrivAer Notchback model, constructed in SolidWorks 2020 using the publicly available 3D data provided by the Technical University of Munich [19].

## 3.2 Computational Domain

Based on the defined problem conditions, the governing equations are solved within a designated region known as the computational domain. To ensure that the boundaries do not influence the simulation results, they must be placed sufficiently far from the vehicle model. As suggested in previous studies [23], the dimensions of this domain are determined in proportion to the vehicle's characteristic length (L), width (W), and ride height (H). Specifically, the inlet boundary is located at a distance of 2L upstream of the front bumper, the outlet boundary at 7L downstream of the rear bumper, while the height and width of the domain are set to 8H and 11W, respectively. In addition to the main domain, a refined subdomain is defined around the vehicle to enhance the accuracy of vortex prediction. This subdomain increases mesh resolution in regions near the body, improving the capture of flow separation phenomena. Within this setup, the front and top boundaries of the subdomain are positioned 30 cm away from the front bumper and the vehicle's roof, respectively, and the rear boundary is placed at a distance of L behind the rear bumper, as illustrated in Figure 2. This spacing is particularly important for accurately resolving the wake region, where flow separation and vortex formation are most prominent.

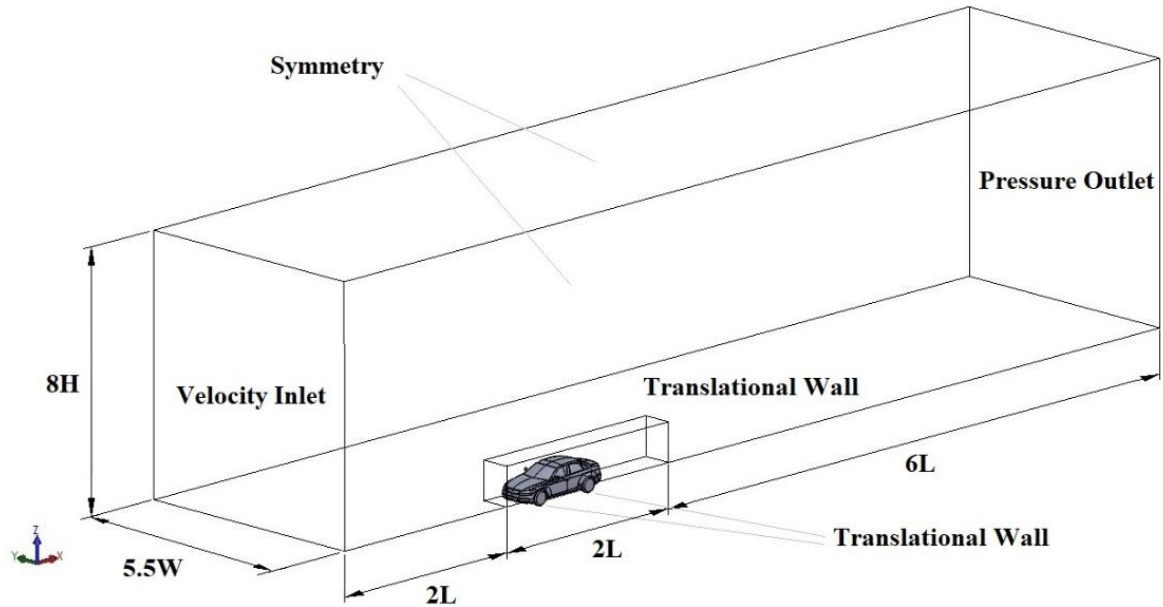

**Figure 2.** The geometry and dimensions of the computational domain, along with the applied boundary conditions

## 3.3 Computational Domain Meshing

The computational domain was meshed using the ANSYS Meshing environment. Due to the geometric complexity of the vehicle body, an unstructured triangular mesh was applied. Furthermore, considering the presence of the viscous sublayer in the boundary layer near the wall and the use of the SST $k - \omega$ turbulence model, a special near-wall mesh satisfying the condition $y^+ \leq 1$ was required in this region [24]. Assuming $y^+ = 1$ [25] and a Reynolds number of $4.87 \times 10^6$, the height of the first cell adjacent to the wall was calculated to be 0.025 mm. Additionally, 10 inflation layers (Figure 3(b)) were defined with a growth rate of 1.2 to accurately resolve the boundary layer. As shown in Figure 3(a), the mesh distribution around the DrivAer Notchback model in the symmetry plane is illustrated. To evaluate the mesh quality in the near-wall regions, the $y^+$ distribution (Figure 3(c)) was plotted. The results indicate that the $y^+$ values successfully meet the target of being close to 1, with a maximum value of approximately 1.3.

(a)

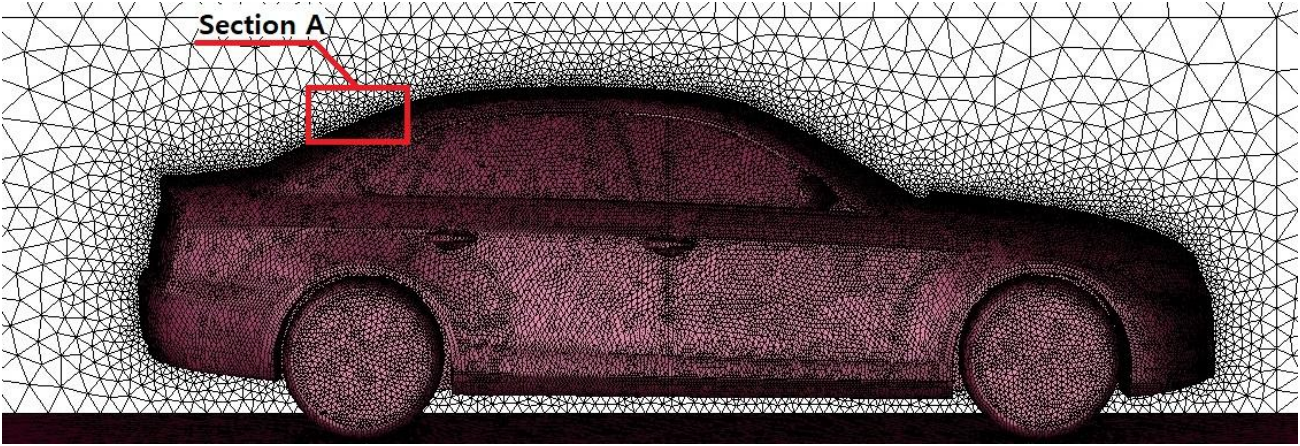

(b)                                    (c)

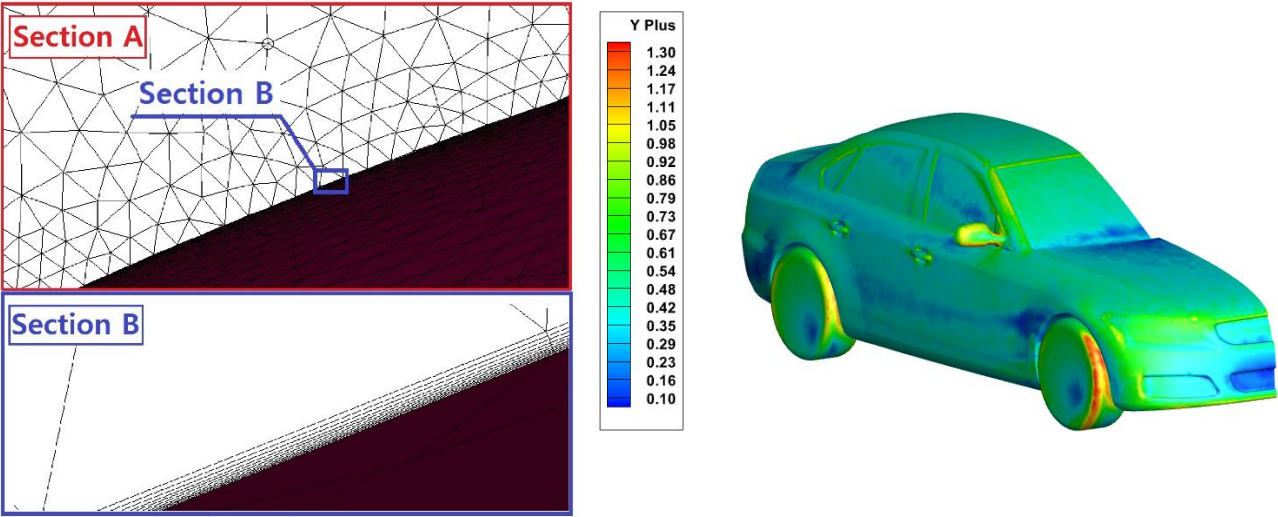

**Figure 3.** (a) The computational meshes generated in the ANSYS Meshing 2019 environment, around the DrivAer Notchback model in the symmetry plane (b) Meshing and inflation details around the body (c) $y^+$ distribution on body and wheels

### 3.4 Boundary Conditions and Numerical Solver Settings

In the aerodynamic evaluation of ground vehicles, air is typically considered an incompressible fluid when the Mach number remains below 0.3, as density variations under such conditions are generally less than 5% and thus negligible for numerical simulations [26, 27]. Accordingly, the air density and dynamic viscosity are assumed constant, with values of 1.225 kg/m³ and $1.7894 \times 10^{-5}$ N·s/m², respectively, consistent with the default settings of ANSYS Fluent [28]. The frontal area of the Notchback model facing the airflow was precisely calculated using the *Silhouette* feature in SolidWorks, yielding a value of 2.185 m². Due to the geometrical symmetry of the vehicle, only half of the computational domain was defined in the simulation to reduce computational cost. The flow analysis was conducted under steady-state conditions using a pressure-based solver. A second-order scheme was adopted for spatial discretization to enhance accuracy, while the SIMPLEC algorithm was employed for coupling pressure and velocity fields. The Green-Gauss Cell-Based method was selected for gradient reconstruction due to its suitability for unstructured tetrahedral meshes and high accuracy in computing derivatives [29]. As shown in Figure 2, the inlet boundary was set as a velocity inlet determined by the Reynolds number (e.g., 15.421 m/s for $Re = 4.87 \times 10^6$), while the outlet was defined as a pressure outlet with a gauge pressure of 0 Pa. A symmetry boundary condition was applied at the vehicle's midplane, and the top and lateral boundaries of the computational domain were assigned zero-gradient conditions for all flow variables to prevent artificial fluxes. The vehicle body was modeled as a stationary wall with a no-slip condition, while the ground was treated as a moving wall with a velocity equal to the freestream to accurately simulate road conditions. Additionally, the wheels were defined as rotating walls with no slip, and the rotational speed was calculated based on the specified Reynolds number and a wheel radius of 0.33 m, resulting in an angular velocity of 46.73 rad/s.

### 3.5 Mesh Independence Study

Before conducting numerical simulations of the airflow around the Audi A4, a mesh independence study was carried out using the DrivAer Notchback model under identical boundary conditions and turbulence modeling approach. The resolution of the mesh significantly affects the accuracy of numerical results: insufficient refinement may lead to inaccuracies, while excessive refinement increases computational cost without meaningful improvements in precision [30]. As illustrated in Figure 4(a), the drag coefficient was monitored as the number of cells increased, with values calculated after achieving convergence for all governing equations. The drag coefficient stabilized at approximately 0.254 beyond 2.34 million cells. This value shows a small deviation of around 3.25% compared to the experimentally obtained value of 0.246 [5, 31]. Such a discrepancy can be attributed to several factors, including the idealized boundary conditions applied in the simulation (e.g., uniform inlet velocity, turbulence intensity assumptions), slight geometric simplifications, and the absence of surface roughness effects in the model. Despite this minor difference, the agreement between the numerical and experimental results is considered satisfactory and confirms the validity of the mesh setup and the selected turbulence model. Moreover, to improve validation, the local pressure coefficient distribution was examined against mesh size. As shown in Figure 5, when the cell count exceeded approximately 2.3 million, the numerical curves (orange and red) aligned more closely with experimental data (black circles). The figure clearly demonstrates the pressure differential between the front and rear sections of the vehicle, highlighting pressure drag as the primary resistance force. Although numerical and experimental results align well in the front regions, noticeable discrepancies remain in the rear sections (refer to Section B in Figure 5), especially around the flow separation zones near the roofline and the notchback-style rear window.

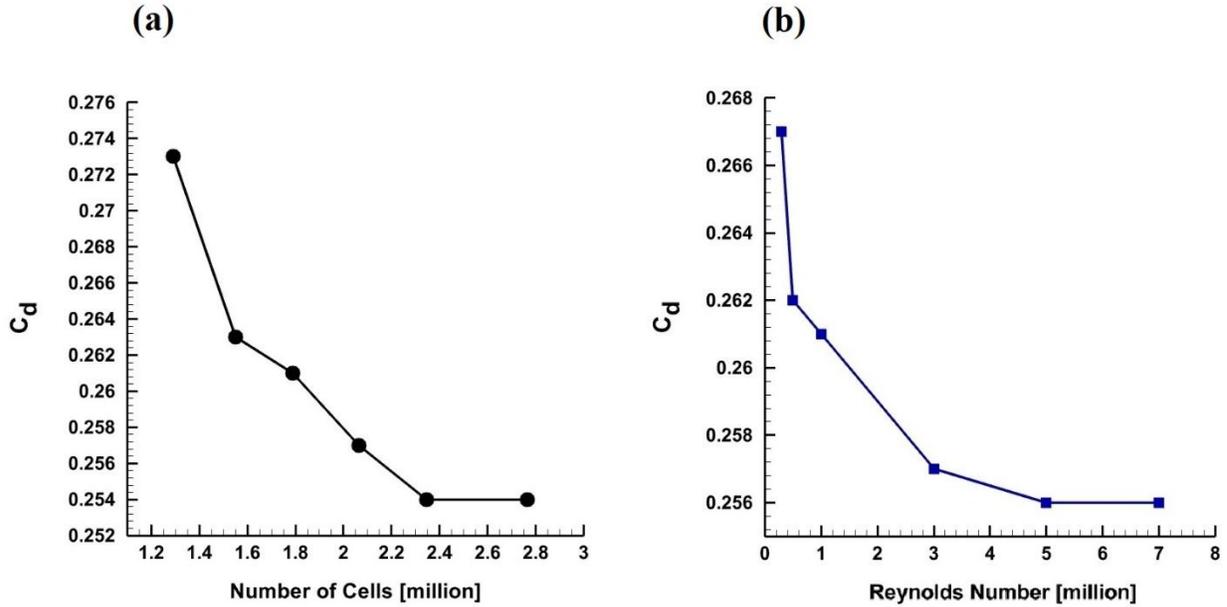

**Figure 4.** Variation of the total drag coefficient for the DrivAer Notchback model with respect to (a) mesh resolution and (b) Reynolds number.

### 3.6 Rationale for Considering Multiple Reynolds Numbers

In aerodynamic simulations, it is well established that the aerodynamic coefficients of a fixed vehicle geometry, such as drag ($C_d$) and lift ($C_l$), tend to stabilize at sufficiently high Reynolds numbers. In this regime, the flow is fully turbulent and largely independent of viscosity-dominated effects, so further increases in Re result in negligible changes in $C_d$ and $C_l$ [22]. For bluff bodies like notchback vehicles, this stabilization typically occurs around Re $\approx 5 \times 10^6$ [32, 33]. For validation purposes, a notchback geometry resembling the real Audi A4 was used [5]. This choice is motivated by the strong similarity of the Audi A4 to the DrivAer model, which allows a reliable comparison between simulation results and available reference data. As shown in Figure 4(b), in this simulation the aerodynamic coefficients of the notchback DrivAer geometry stabilize and become nearly independent of Reynolds number at approximately Re $\approx 5 \times 10^6$.

It is important to distinguish between Reynolds number independence of a fixed geometry and Reynolds number robustness of an optimized geometry. While a standard notchback vehicle may exhibit stable aerodynamic coefficients beyond $Re \approx 5 \times 10^6$, there is no a priori guarantee that the shape obtained from the optimization process will perform equally consistently across different Reynolds numbers. The purpose of testing multiple Reynolds numbers in this study is therefore not merely to demonstrate the stability of the standard geometry, but to examine the key hypothesis that the optimized vehicle geometry remains robust against variations in Re. This approach ensures that the optimized design maintains stable aerodynamic performance under realistic operating conditions and justifies the computational effort of running simulations at multiple Reynolds numbers.

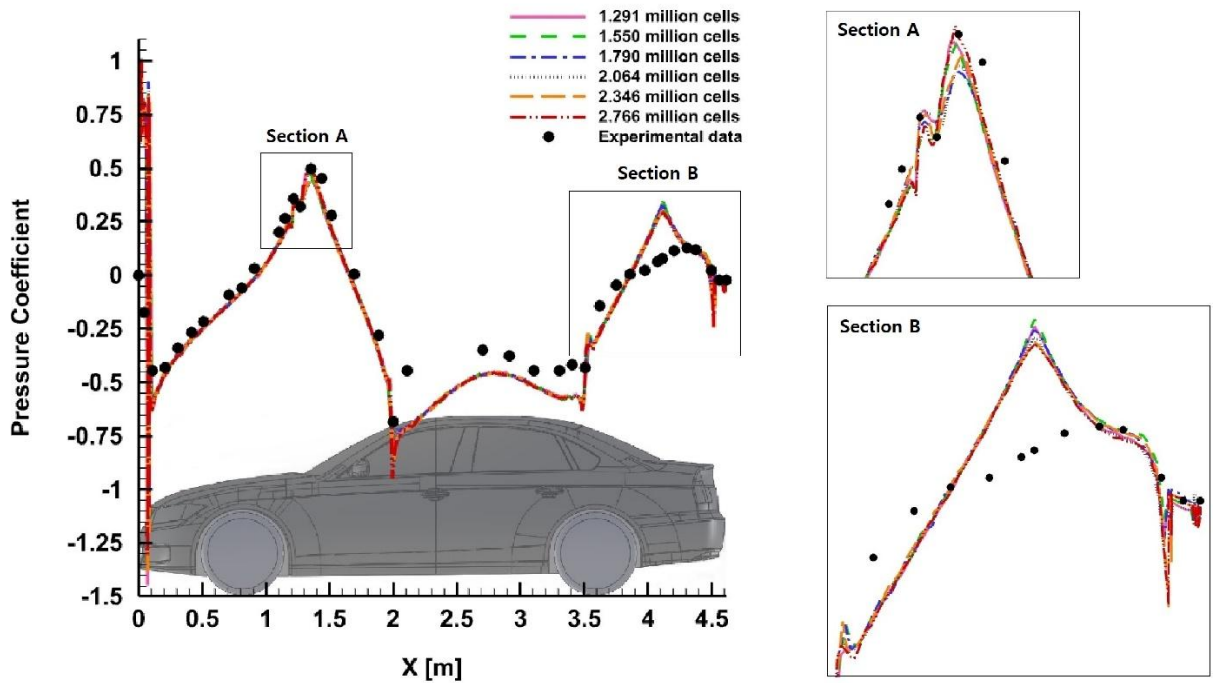

**Figure 5.** Distribution of the static pressure coefficient over the upper surface of the DrivAer Notchback model for different mesh resolutions, compared with experimental wind tunnel data [5] along the centerline.

## 4. Results and Discussion

In this study, the optimal ride height and rake angle for the Audi A4 vehicle under various flow conditions with different Reynolds numbers are determined using a machine learning algorithm, with the aim of achieving desirable aerodynamic performance. Under normal driving conditions, optimal aerodynamic behavior occurs when the drag ($C_d$) and lift ($C_l$) coefficients reach their minimum values, leading to lower fuel consumption and enhanced driving stability [34]. However, in special cases such as slippery roads or racing scenarios, the desirable aerodynamic conditions differ and must be adjusted accordingly. Since the vehicle's posture constantly changes while driving, any adjustable system used to modify ride height and rake angle must be equipped with an adaptive control strategy to adjust in real-time based on varying conditions such as speed [35]. Nevertheless, simulating all aerodynamic conditions across a continuous design domain is computationally expensive. Therefore, to train the machine learning algorithm, a dataset of drag and lift coefficients was generated through simulations within a relatively broad range of ride heights and rake angles, under four flow conditions with Reynolds numbers of $4.87\times10^6$, $9.75\times10^6$, $14.61\times10^6$, and $19.48\times10^6$. Ultimately, the trained AI model is capable of accurately predicting the optimal ride height and rake angle under the desired aerodynamic performance criteria.

### 4.1 Definition of the Range of Ride Height and Rake Angle for the Vehicle

The range of ride height and rake angle variations must be carefully selected to ensure that it covers sufficient design scenarios while avoiding excessive deformation or increased design and implementation costs. Furthermore, the ride height and rake angle must not result in a ground clearance lower than the minimum safe distance, to prevent any collision between the vehicle's underbody components and road bumps or irregularities. According to previous studies, for racing cars, a ground clearance of less than 80 mm is not permitted. However, in this study, the minimum ground clearance was set to 100 mm, providing a balanced compromise between the racing standard and the general minimum requirement of 120 mm for passenger vehicles [36, 37]. The vehicle ride height was varied from 1.336 m to 1.536 m with a ±10 cm range from the baseline and 5 cm intervals. The rake angle was varied from 0° to 5° in 1° increments. Figures 6(a) to 6(e) illustrate a schematic of the vehicle under five different ride heights at 0° rake angle, while Figures 6(f) to 6(j) show a schematic of the vehicle at the baseline ride height of 1.436 m under six different rake angles.

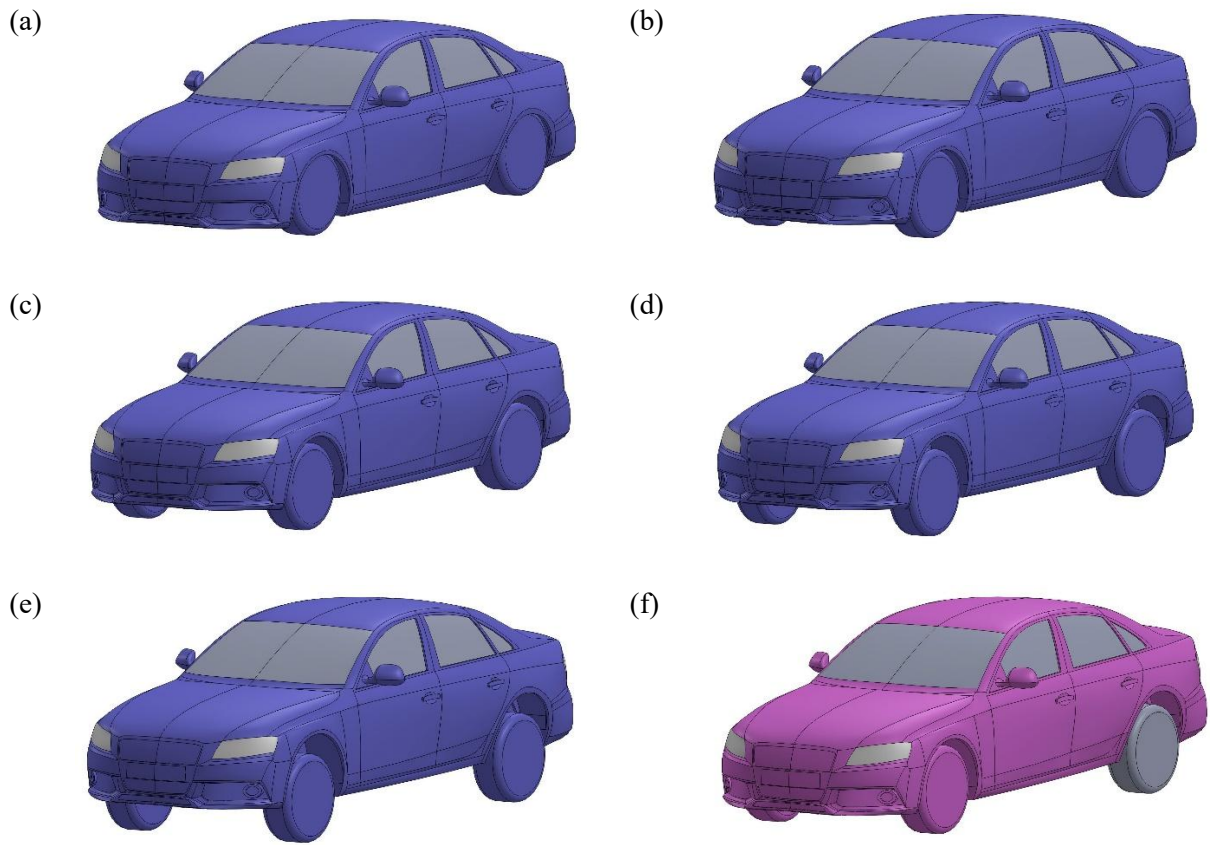

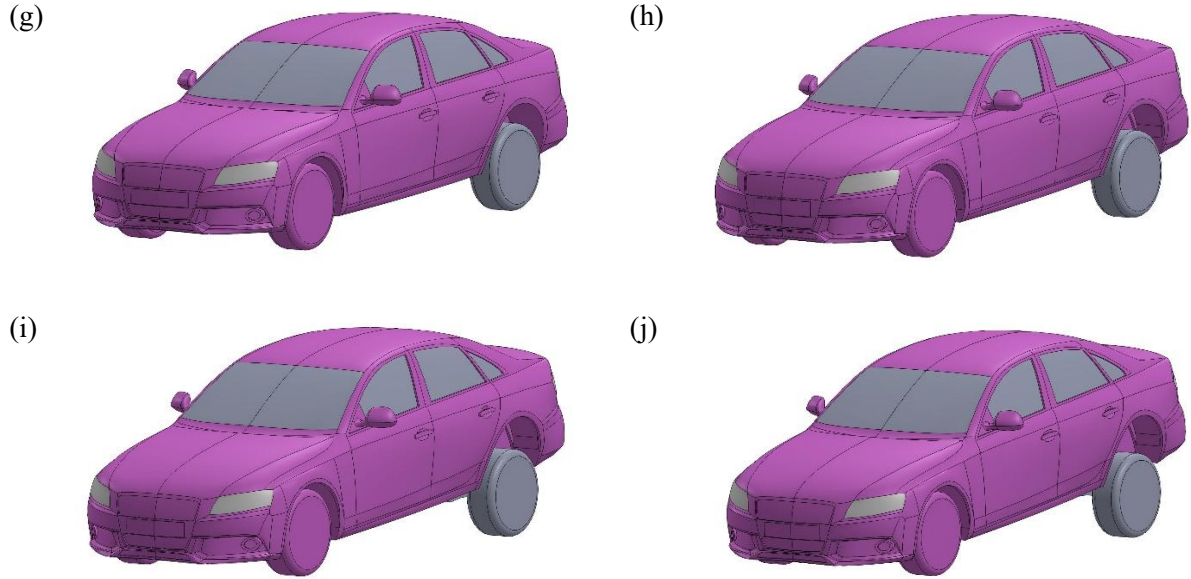

(g)      (h)

(i)      (j)

**Figure 6.** Three-dimensional schematic of the Audi A4 vehicle at zero rake angle under five different ride heights: a) 1.336 m, b) 1.386 m, c) 1.436 m, d) 1.486 m, and e) 1.536 m; and at the baseline ride height of 1.436 m under various rake angles: f) 1°, g) 2°, h) 3°, i) 4°, and j) 5°.

## 4.2 Effect of Ride Height Variation at Zero Rake Angle

As shown in Figures 6(a) to 6(e), the three-dimensional geometry of the Audi A4 vehicle is illustrated at various overall ride heights ranging from 1.336 meters to 1.536 meters in 5-centimeter increments, all at a rake angle of zero degrees. The baseline geometry corresponds to a height of 1.436 meters with a rake angle of zero degrees. In this section, the rake angle is kept constant while the effect of varying ride height on the aerodynamic coefficients is analyzed. The results of the simulations for different ride heights are presented in Figure 7, illustrating how the drag and lift coefficients change with increasing or decreasing ride height at a given Reynolds number.

**(a)**                    **(b)**

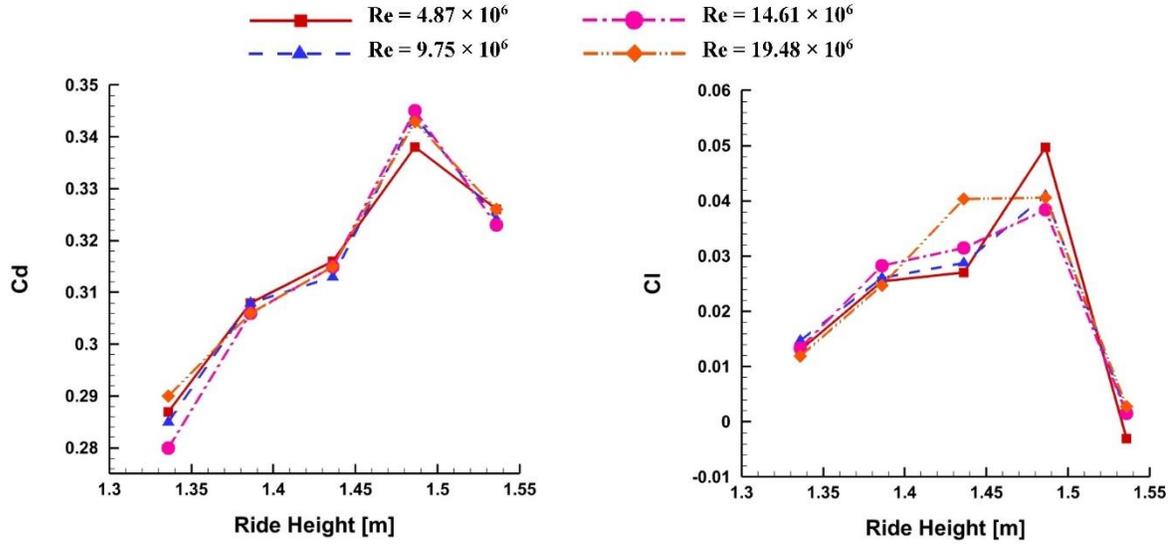

**Figure 7.** The variations of (a) drag and (b) lift coefficients at different ride heights at a given Reynolds number.

As shown in Figure 7(a) and 7(b), the variations in drag coefficient and lift coefficient for the Audi A4 model with different ride heights at a fixed rake angle of 0° have been presented. The ride height was varied from 1.336 m to 1.536 m in 5 cm steps, covering four different Reynolds numbers: $4.87 \times 10^6$, $9.75 \times 10^6$, $14.61 \times 10^6$, and $19.48 \times 10^6$. The results clearly demonstrate that both aerodynamic coefficients are significantly affected by changes in ride height, and the influence varies depending on the Reynolds number. At each Reynolds number, the drag coefficient decreases as the vehicle height is lowered to 1.336 m and then increases again as the height is raised beyond the baseline value of 1.436 m. This trend indicates that the minimum drag occurs at the lowest tested ride height of 1.336 m, highlighting the aerodynamic benefit of reducing ride height. For instance, at $Re = 4.87 \times 10^6$, $C_d$ drops from 0.316 (baseline ride height) to 0.287 at 1.336 m, then rises to 0.338 and 0.326 at 1.486 m and 1.536 m respectively. As the Reynolds number increases, the drag coefficients generally remain in a similar trend, but the magnitude of reduction and increase becomes more pronounced. At $Re = 14.61 \times 10^6$, the lowest $C_d$ is 0.280 at 1.336 m, while at 1.486 m it rises sharply to 0.345, the highest observed drag in the dataset. This suggests that at higher speeds (larger Reynolds numbers), the aerodynamic penalties for increased ride height are more severe. In terms of lift coefficient, a comparable pattern is observed. $C_l$ values increase as height is reduced to 1.336 m, but rise even more at 1.486 m before slightly dropping again at the maximum height. However, in almost all cases, the lift coefficient remains positive, especially at mid and lower ride heights, indicating a net upward force. Notably, only at 1.536 m and $Re = 4.87 \times 10^6$ does the $C_l$ become negative ($-0.003$), implying a marginal net downforce. Overall, this analysis shows that reducing ride height improves aerodynamic efficiency by lowering drag, though it can also slightly increase lift in most cases. The influence of Reynolds number becomes more significant at higher values, magnifying both drag penalties and lift tendencies. Thus, careful optimization of ride height is essential for achieving a balanced aerodynamic performance, especially at high speeds.

### 4.3 Effect of Rake Angle Variation at the Baseline Ride Height

As illustrated in Figures 6(f) to 6(j), the vehicle rake angle is varied from 0 to 5 degrees in 1-degree increments. Given that simulations were conducted across five different ride heights and six different rake angles, and considering that this section aims to demonstrate the overall effect of rake angle as a single variable, the analysis presented here is limited to the baseline ride height of 1.436 meters, the height at which the vehicle was originally designed. As shown in Figure 8, since the rake angle change is implemented by adjusting the hydraulic jacks at the rear (with rotation around the front axle), it should be noted that at lower ride heights (e.g., 1.336 meters), increasing the rake angle causes the vehicle's nose to come too close to the ground. Therefore, to ensure the minimum ground clearance of 100 mm is maintained, only the 1-degree rake angle was simulated for ride height of 1.336 meters. For higher rake angles, the clearance drops below the allowable threshold, making those simulations invalid under the defined constraints. However, for all other tested ride heights, no such limitations existed, and the full range of rake angles was simulated.

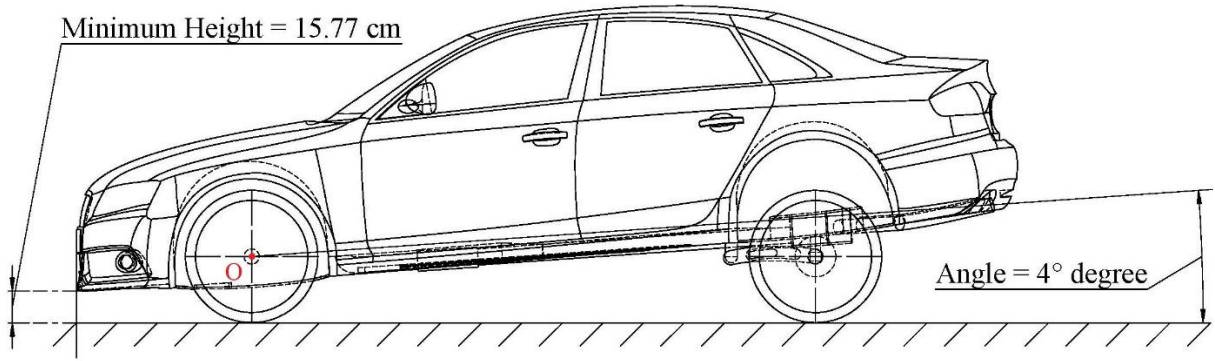

**Figure 8.** A depiction of the rake angle position and the minimum ground clearance beneath the vehicle for a sample case with a 4-degree rake angle and an ride height of 1.436 meters

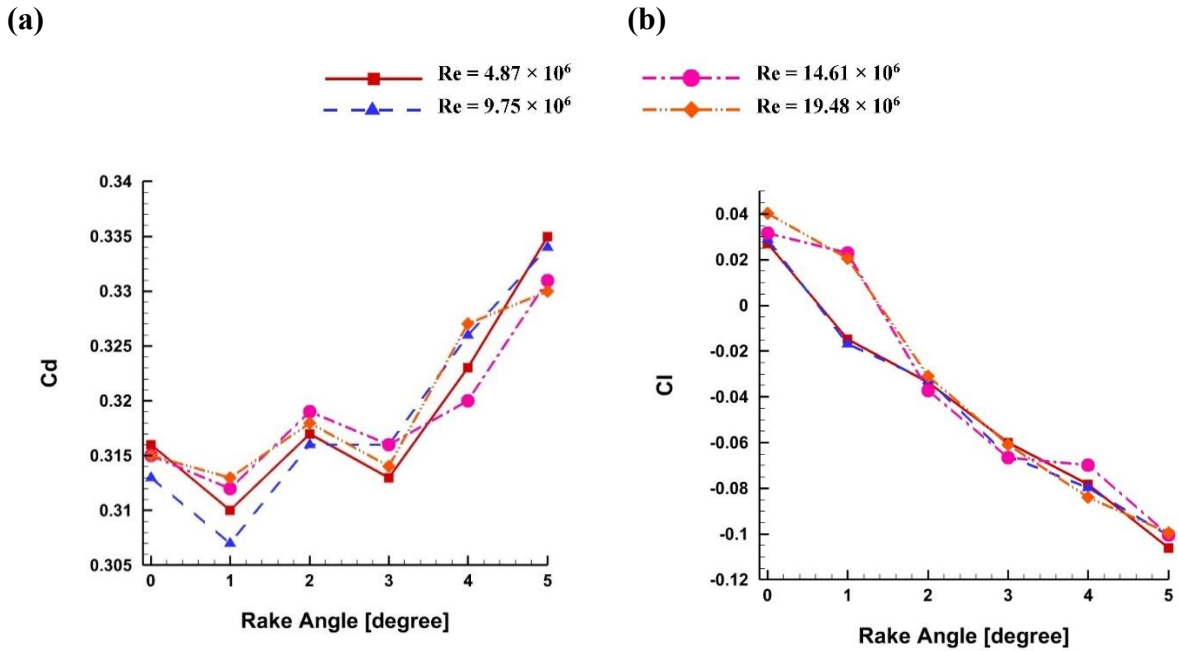

**Figure 9.** Variations of (a) drag and (b) lift coefficients at an ride height of 1.436 meters and at different rake angles at a given Reynolds number.

In the second part of this study, the combined effects of Reynolds number and rake angle on the aerodynamic drag and lift coefficients were examined under the condition of a fixed ride height of 1.436 meters. This height corresponds to the baseline geometry before any change in rake angle was applied. All data analyzed in this section are extracted solely for this constant height. The results are illustrated in Figures 9(a) ($C_d$) and 9(b) ($C_l$). As shown in Figure 9(a), increasing the rake angle from 1° to 5° generally leads to a rise in the drag coefficient. At the lowest Reynolds number ($4.87 \times 10^6$), the drag increases from 0.310 at 1° to 0.335 at 5°. A similar trend is observed across higher Reynolds numbers; for instance, at $19.48 \times 10^6$, the drag increases from 0.313 to 0.330 over the same angle range. This increase is primarily attributed to a larger frontal inclination of the vehicle, which leads to expanded flow separation regions and increased pressure drag. Figure 9(b) illustrates that the lift coefficient becomes progressively more negative with increasing rake angle. This indicates a significant enhancement of the downward aerodynamic force (downforce). At 1°, the lift coefficient at low Reynolds number is -0.0149, whereas it drops to -0.1061 at 5°. A similar pattern is observed at higher Reynolds numbers; for example, at $19.48 \times 10^6$, the lift coefficient decreases from 0.0206 at 1° to -0.0996 at 5°. This behavior suggests that the rake angle promotes upper surface suction and lower surface pressure, both contributing to stronger downforce. In summary, for a constant ride height of 1.436 meters, increasing the rake angle consistently results in higher drag and stronger downforce across all Reynolds numbers studied. These findings highlight the aerodynamic trade-offs associated with geometric adjustments and emphasize the importance of simultaneous optimization of both drag and lift in performance-oriented vehicle design.

## 4.4 Machine Learning Framework for Aerodynamic Prediction

### 4.4.1 Model Development and Evaluation Strategy

Ground-vehicle aerodynamics is strongly governed by ride height, rake angle, and Reynolds number. While prior studies often examined these effects in isolation, assessing their combined influence is nontrivial and cannot rely solely on simulations [1, 2]. Therefore, advanced machine learning (ML) algorithms were employed to construct predictive models over a broad operating envelope. Three state-of-the-art algorithms were evaluated: Random Forest, Gradient Boosting, and LightGBM [38–40]. Models were trained, validated, and tested on the prepared dataset, and assessed using R², RMSE, MAE, and MAPE [41, 42] for both drag ($C_d$) and lift ($C_l$).

### 4.4.2 Parity Analysis of Predictions

Parity plots (Figure 10) illustrate agreement between predictions and ground truth; proximity to the 45° line indicates higher accuracy. Results show that Gradient Boosting consistently provides the closest alignment with the reference line for both $C_d$ and $C_l$, confirming its superior predictive accuracy compared to Random Forest and LightGBM [43].

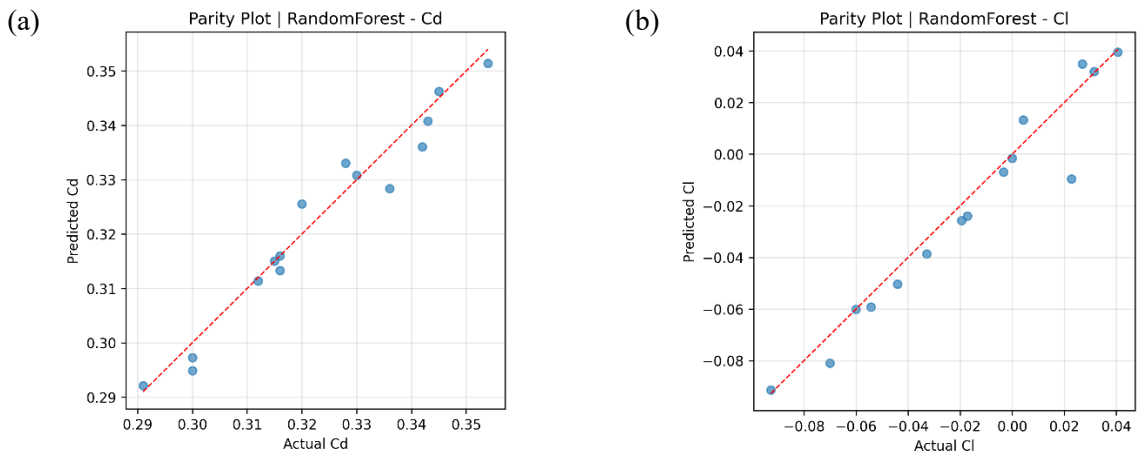

(a) Parity Plot | RandomForest - Cd

(b) Parity Plot | RandomForest - Cl

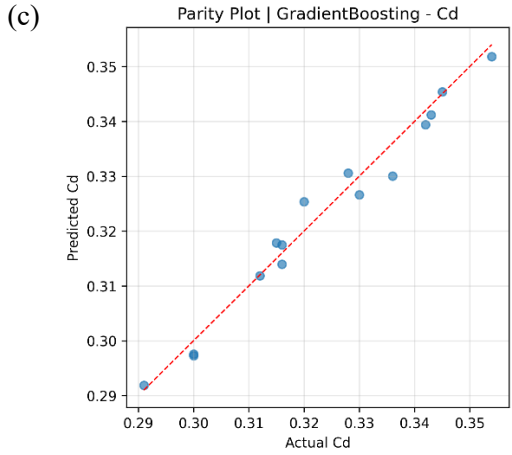
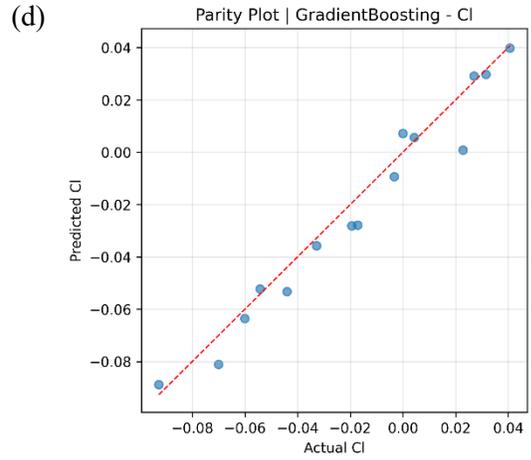

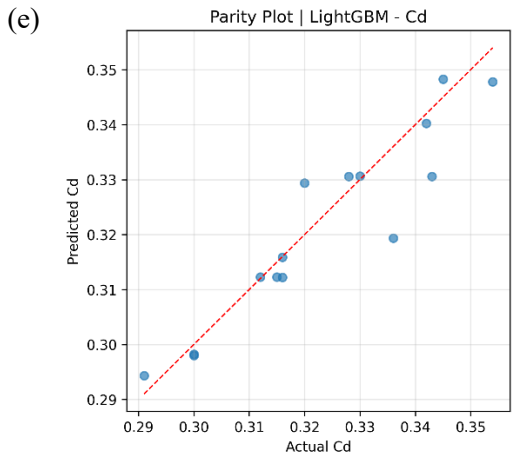
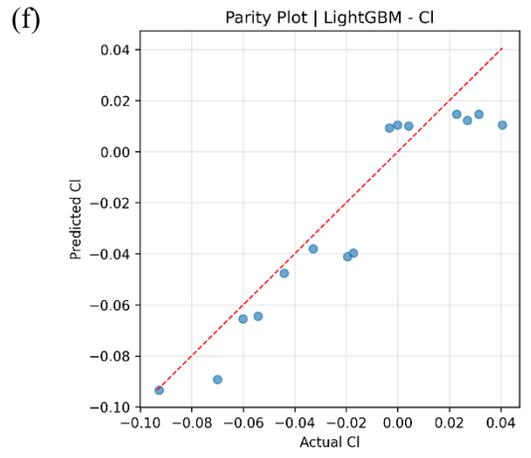

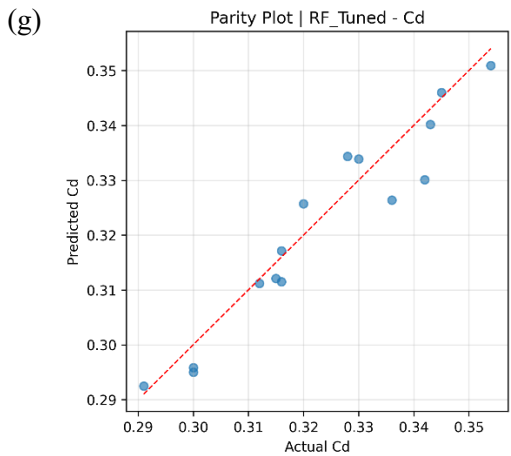
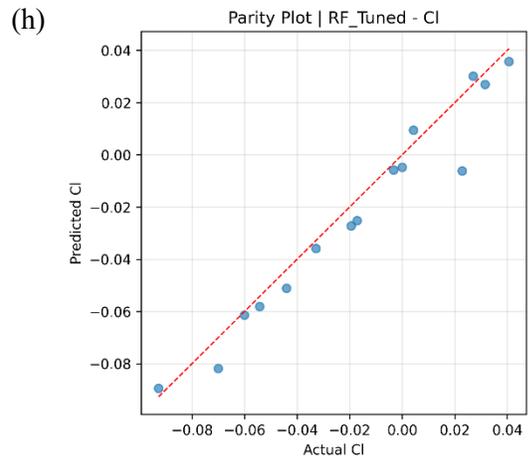

**Figure 10.** Parity plots of actual vs. predicted aerodynamic coefficients for RF (a) $C_d$, (b) $C_l$, XGBoost (c) $C_d$, (d) $C_l$, and LightGBM (e) $C_d$, (f) Cl, RF_Tuned (g) $C_d$, (h) $C_l$

### 4.4.3 Quantitative Performance Metrics

Table 1 summarizes the metrics for the tested models. Gradient Boosting achieved the best overall results, with $R^2$ values of 0.9739 ($C_d$) and 0.9555 ($C_l$), accompanied by the lowest error values across metrics. Random Forest and its tuned variant demonstrated competitive performance, particularly for $C_l$ prediction, whereas LightGBM yielded weaker accuracy for both targets. These findings demonstrate the robustness of Gradient Boosting in capturing the nonlinear aerodynamic dependencies [44]. It should be noted that the reported MAPE values for $C_l$ are relatively large (e.g., 519% for Gradient Boosting). This behavior originates from the fact that $C_l$ values are close to zero in several cases, making the MAPE metric unstable and less informative. Therefore, $R^2$ and RMSE provide a more reliable assessment of the predictive accuracy for $C_l$.

**Table 1.** Quantitative comparison of ML models for $C_d$ and $C_l$

| Model | Target | R2 | RMSE | MAE | MAPE (%) |
|---|---|---|---|---|---|
| RandomForest | $C_d$ | 0.9571 | 0.003714 | 0.0029 | 0.891719 |
| RandomForest | $C_l$ | 0.9337 | 0.010054 | 0.006566 | 137.6266 |
| GradientBoosting | $C_d$ | 0.9739 | 0.002899 | 0.002457 | 0.758479 |
| GradientBoosting | $C_l$ | 0.9555 | 0.008243 | 0.006218 | 519.6587 |
| LightGBM | $C_d$ | 0.8717 | 0.006421 | 0.004462 | 1.353282 |
| LightGBM | $C_l$ | 0.856 | 0.01482 | 0.012489 | 776.6212 |
| RF_Tuned | $C_d$ | 0.9136 | 0.005269 | 0.004293 | 1.317332 |
| RF_Tuned | $C_l$ | 0.9432 | 0.00931 | 0.006659 | 347.6916 |

## 4.5 Model Interpretability and Optimization Framework

### 4.5.1 Dataset and Splitting Strategy

The dataset included three inputs (Ride Height, Rake Angle, Re) and two outputs ($C_d$, $C_l$). After preprocessing (removing NaNs and constraining rake angle to 0–5°), the data were split into 70%/15%/15% (train/validation/test) using a fixed random seed (42) to ensure reproducibility [45].

### 4.5.2 Selected Models and Hyperparameter Tuning

To illustrate the implemented workflow, a schematic diagram is provided (Figure 11), outlining the applied machine learning algorithms, including Random Forest, Gradient Boosting, and LightGBM. The diagram highlights the overall pipeline of training, validation, testing, and hyperparameter tuning. Given the superior test results, Gradient Boosting was selected for both $C_d$ and $C_l$ predictions. Hyperparameter tuning for Random Forest was also conducted to ensure fairness of comparison, with the best parameters reported in Table 2. However, the Gradient Boosting regressors consistently outperformed all alternatives and were therefore adopted for subsequent optimization analysis [46].

**Table 2.** Optimal hyperparameters for the RF_Tuned model, reported for comparison with Gradient Boosting

| Target | n_estimators | max_depth | min_samples_split | min_samples_leaf |
|--------|--------------|-----------|-------------------|------------------|
| $C_d$  | 141          | 23        | 4                 | 1                |
| $C_l$  | 147          | 25        | 4                 | 1                |

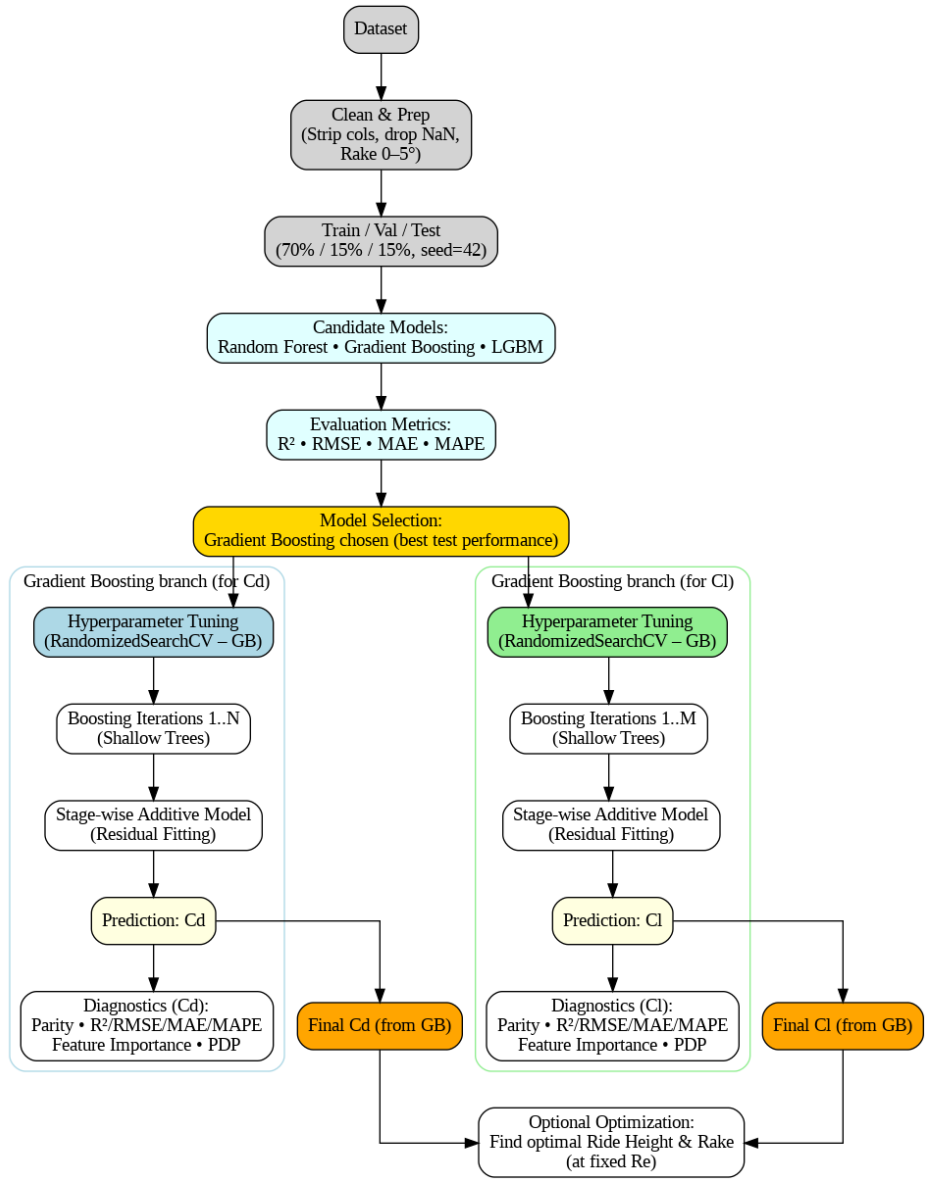

**Figure 11.** Schematic of the machine learning workflow and model selection process.

### 4.5.3 Feature Importance Analysis

Feature-importance results (Figure 12) indicate that ride height is the dominant parameter influencing $C_d$, followed by rake angle and Re. In contrast, $C_l$ is primarily governed by rake angle, with ride height as the secondary factor and Re showing minimal contribution. These interpretations are consistent with aerodynamic theory and prior CFD-based studies [22, 47].

(a) (b)

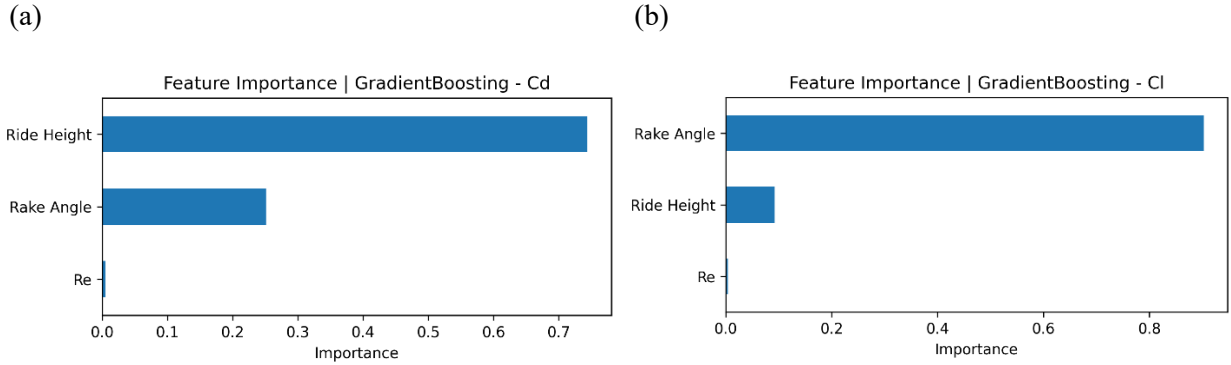

**Figure 12.** Feature importance from Gradient Boosting: (a) $C_d$, dominated by ride height; (b) $C_l$, dominated by rake angle.

### 4.5.4 Partial Dependence Plots (PDP)

The partial dependence plots in Figure 13(a) demonstrate that ride height has a pronounced nonlinear effect on the drag coefficient. While lowering the ride height initially reduces drag, $C_d$ increases sharply beyond a certain threshold due to underbody flow restriction. The influence of rake angle on $C_d$ is monotonic but comparatively milder. Conversely, Figures 13(c) and 13(d) show that rake angle plays the dominant role in determining the lift coefficient; increasing rake enhances downforce until approaching saturation. Ride height exerts a nonlinear secondary effect on $C_l$, whereas Reynolds number contributes negligibly in both cases. These findings are consistent with aerodynamic fundamentals and previously reported CFD-based studies [48, 5].

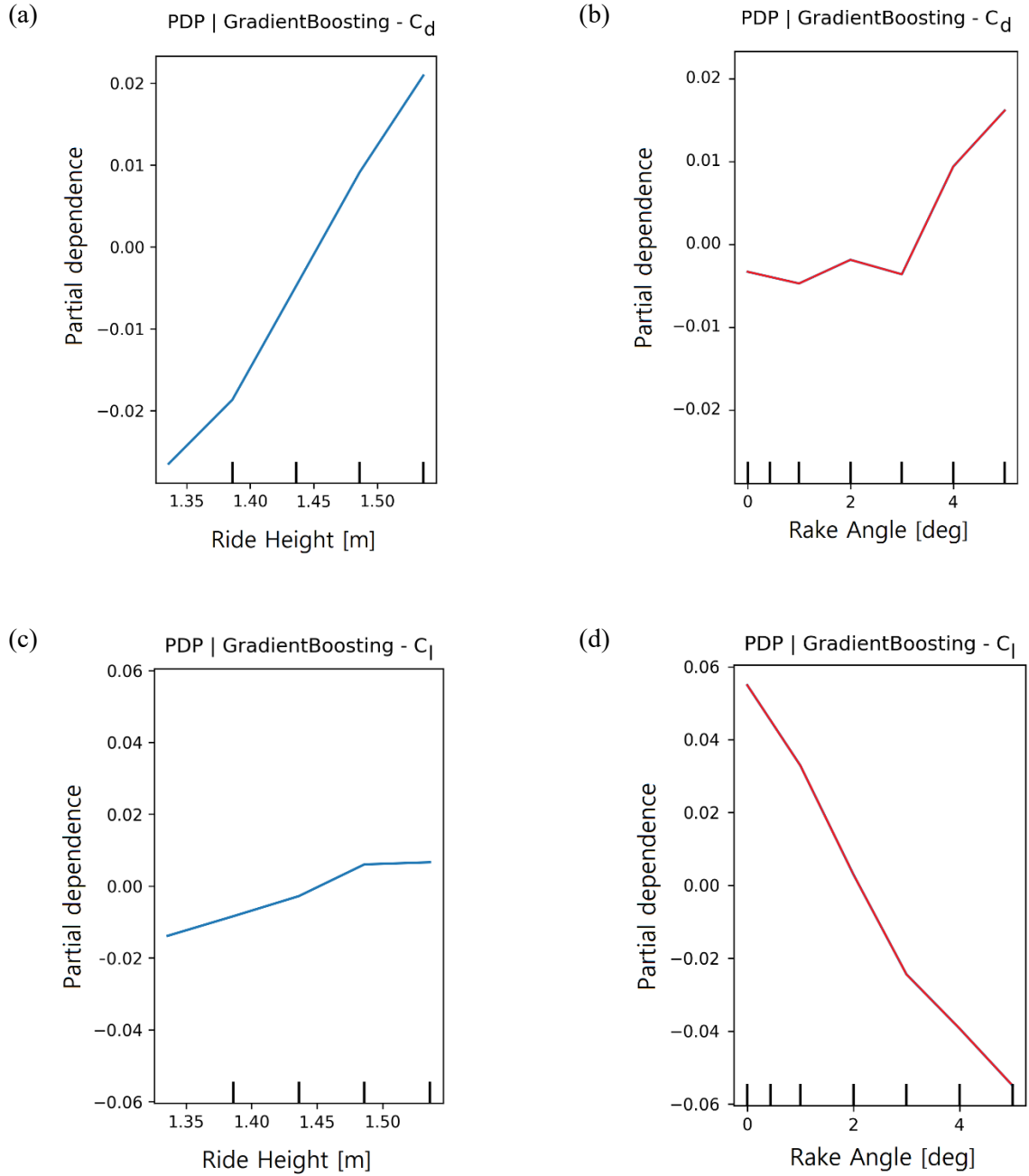

**Figure 13**. Partial dependence plots from Gradient Boosting models: (a) $C_d$ vs ride height, (b) $C_d$ vs rake angle, (c) $C_l$ vs ride height, (d) $C_l$ vs rake angle.

The updated analysis confirms that Gradient Boosting is the most effective model for predicting both $C_d$ and $C_l$. Feature-importance and PDPs jointly demonstrate that $C_d$ is primarily governed by ride height, whereas $C_l$ is chiefly controlled by rake angle; Re has limited influence within the studied range. The optimization results further underline the potential of ML-based surrogate modeling to guide aerodynamic design decisions. Overall, the Gradient Boosting framework yields a reliable, interpretable, and accurate tool for aerodynamic performance prediction and optimization of ground vehicles [49–53].

## 4.6 Optimization of the Combined Effects of Ride Height and Rake Angle on Aerodynamic Parameters Across Reynolds Numbers in the Balanced Condition

To model the relationship between drag ($C_d$) and lift ($C_l$) coefficients with the design variables—ride height, rake angle, and Reynolds number—several machine learning algorithms were tested, including Random Forest Regressor, LightGBM, and Gradient Boosting Regressor. Among them, Gradient Boosting demonstrated the highest accuracy in predictions, better capability in modeling nonlinear dependencies, and stronger generalization, making it the preferred method for subsequent analyses [39, 54]. The models were implemented in Python using the Scikit-learn and LightGBM libraries [45, 40]. Hyperparameters were tuned through RandomizedSearchCV with cross-validation, which proved more efficient than Grid Search for large parameter spaces [55].

The objective function was defined as a weighted combination of $C_d$ and $C_l$:

$$Objective = \alpha \cdot C_d + \beta \cdot C_l \tag{9}$$

In multi-objective optimization problems, it is rarely possible to minimize two conflicting objectives simultaneously; instead, a set of optimal trade-offs, known as the Pareto front, is obtained [7]. Each point on this front represents a solution where improving one objective would necessarily deteriorate the other. In this study, different weighting conditions were not intended as arbitrary choices, but rather as a means of exploring and illustrating representative regions along the Pareto front that balance aerodynamic drag and lift.

It was previously stated that the two parameters, drag coefficient and lift coefficient, play a decisive role in the aerodynamic conditions of vehicles. Accordingly, three different conditions were considered to design and optimize the vehicle's behavior under various circumstances. The first case, referred to as Balanced Condition, was defined in situations where both fuel consumption and vehicle stability during motion are important. Since drag reduction carries greater importance in aerodynamic design, the weighting factors were set to $\alpha = 0.7$ and $\beta = 0.3$. The second condition, known as Minimizing Drag Coefficient, aimed to achieve the lowest aerodynamic resistance and, consequently, minimize fuel consumption. Therefore, the coefficients were determined such that the drag coefficient dominated, i.e., $\alpha = 0.999$ and $\beta = 0.001$. Finally, the third condition, called Maximizing Negative Lift Coefficient, examined conditions where maximum vehicle stability during motion is desired. For this purpose, the coefficients were adjusted as $\alpha = 0.001$ and $\beta = 0.999$.

In this section, the focus is placed on the first case, namely the Balanced Condition. To simultaneously optimize the ride height and rake angle in four Reynolds numbers, the Differential Evolution (DE) algorithm was employed. This population-based optimization method does not require derivative information and is highly resistant to local optima in high-dimensional search spaces [49, 56]. The results of the optimization were presented using contour plots in the design variable space, which clearly highlight the optimal regions for each Reynolds number. Nevertheless, it should be acknowledged that the weighted-sum scalarization approach has inherent limitations. While DE proved effective, scalarization only yields discrete points of the trade-off curve rather than the complete spectrum of Pareto-optimal solutions. Future work may consider adopting multi-objective evolutionary algorithms (MOEAs), such as NSGA-II or MOPSO, to capture the entire Pareto front and provide a more comprehensive design tool [57–59].

(a)
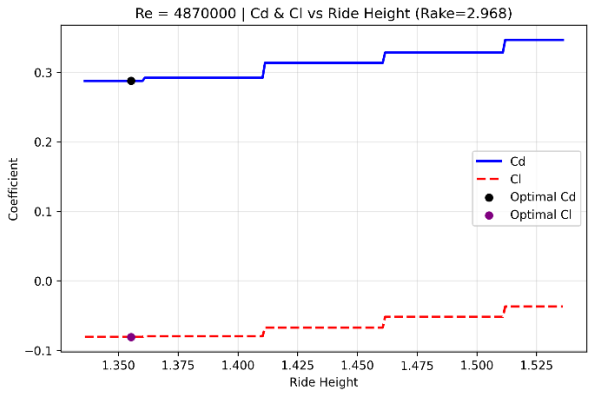

(b)
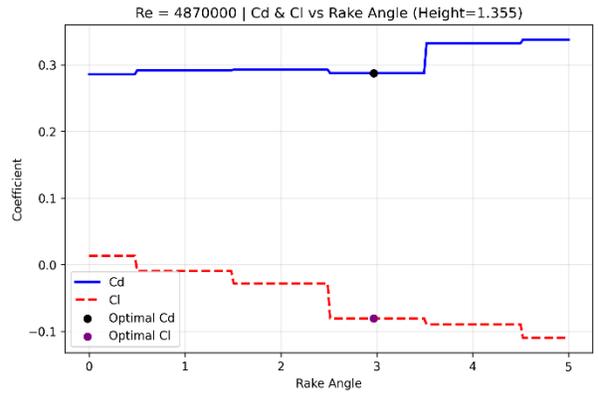

(c)
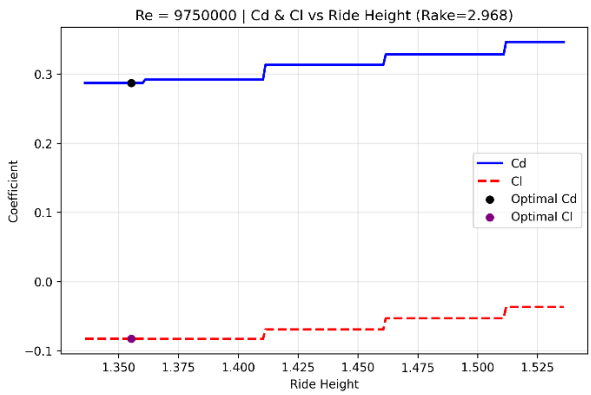

(d)
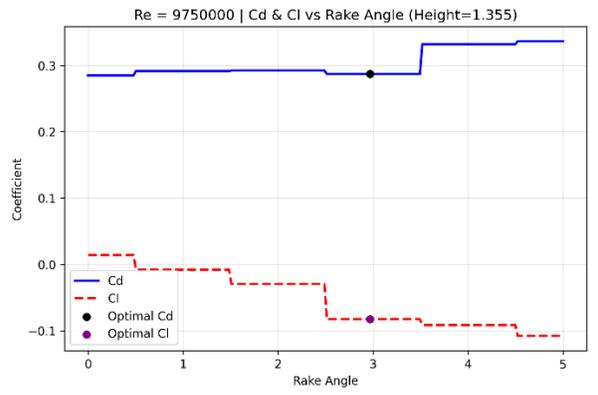

(e)
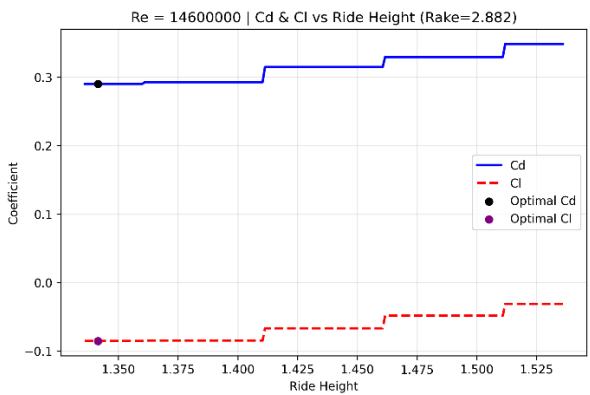

(f)
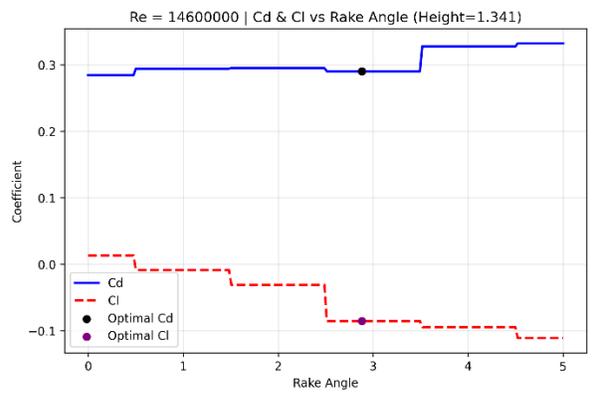

(g)

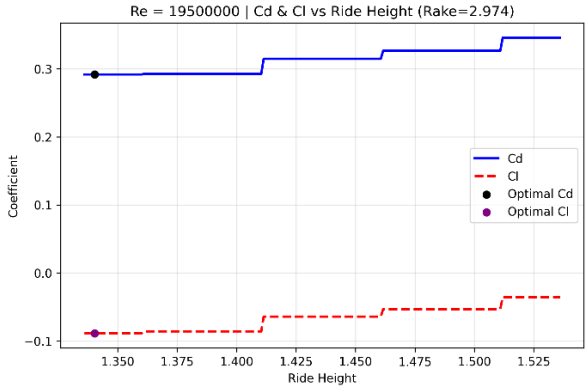

(h)

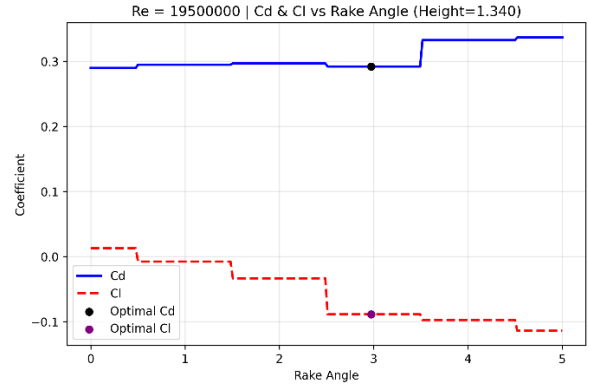

**Figure 14**. Plots of $C_d$ and $C_l$, respectively, based on the combined objective function $(0.7 \cdot C_d + 0.3 \cdot C_l)$ for the Audi A4 at four Reynolds numbers: (a, b) Re = $4.87 \times 10^6$, (c, d) Re = $9.75 \times 10^6$, (e, f) Re = $14.61 \times 10^6$, and (g, h) Re = $19.48 \times 10^6$. For each Reynolds number, the first subfigure [(a), (c), (e), (g)] shows the variation of $C_d$ and $C_l$ with ride height at the optimal rake angle, and the second subfigure [(b), (d), (f), (h)] shows the variation of $C_d$ and $C_l$ with rake angle at the optimal ride height. Optimal points are marked on each curve

Based on the Gradient Boosting optimization results for $C_d$ and $C_l$ at four Reynolds numbers of $4.87 \times 10^6$, $9.75 \times 10^6$, $14.61 \times 10^6$, and $19.48 \times 10^6$ for the Audi A4 model, the optimal ride height and rake angle are clearly identified. Figures 14(a)–14(h) illustrate the variation of $C_d$ and $C_l$ with ride height and rake angle, respectively, with the optimal points marked on each curve. At Re = $4.87 \times 10^6$, the optimal configuration corresponds to a ride height of 1.3554 m and a rake angle of 2.9681°, resulting in a drag coefficient of 0.28751 and a lift coefficient of -0.08062. The weighted objective function $(0.7 \cdot C_d + 0.3 \cdot C_l)$ is 0.17707, indicating a favorable aerodynamic balance between drag reduction and downforce. For Re = $9.75 \times 10^6$, the optimal ride height remains 1.3554 m, while the rake angle stays at 2.9681°, producing $C_d$ = 0.28704, $C_l$ = -0.08264, and an objective value of 0.17614, demonstrating consistent aerodynamic performance with minimal geometric adjustment. At Re = $14.61 \times 10^6$, the ride height slightly decreases to 1.3415 m, with a rake angle of 2.8823°, yielding $C_d$ = 0.29035, $C_l$ = -0.08533, and an objective value of 0.17765, representing a near-optimal balance of drag and downforce in this Reynolds number regime. Finally, at Re = $19.48 \times 10^6$, the ride height is 1.3403 m and the rake angle is 2.9741°, resulting in $C_d$ = 0.29178, $C_l$

= -0.08878, and an objective value of 0.17761, indicating that aerodynamic efficiency is maintained even at high Reynolds numbers, with the optimal configuration confined to a narrow geometric range. Overall, the results reveal that the optimal ride height and rake angle vary nonlinearly with Reynolds number but remain within a compact design space, suggesting that a limited range of geometric settings can sustain near-optimal aerodynamic performance across varied flow conditions.

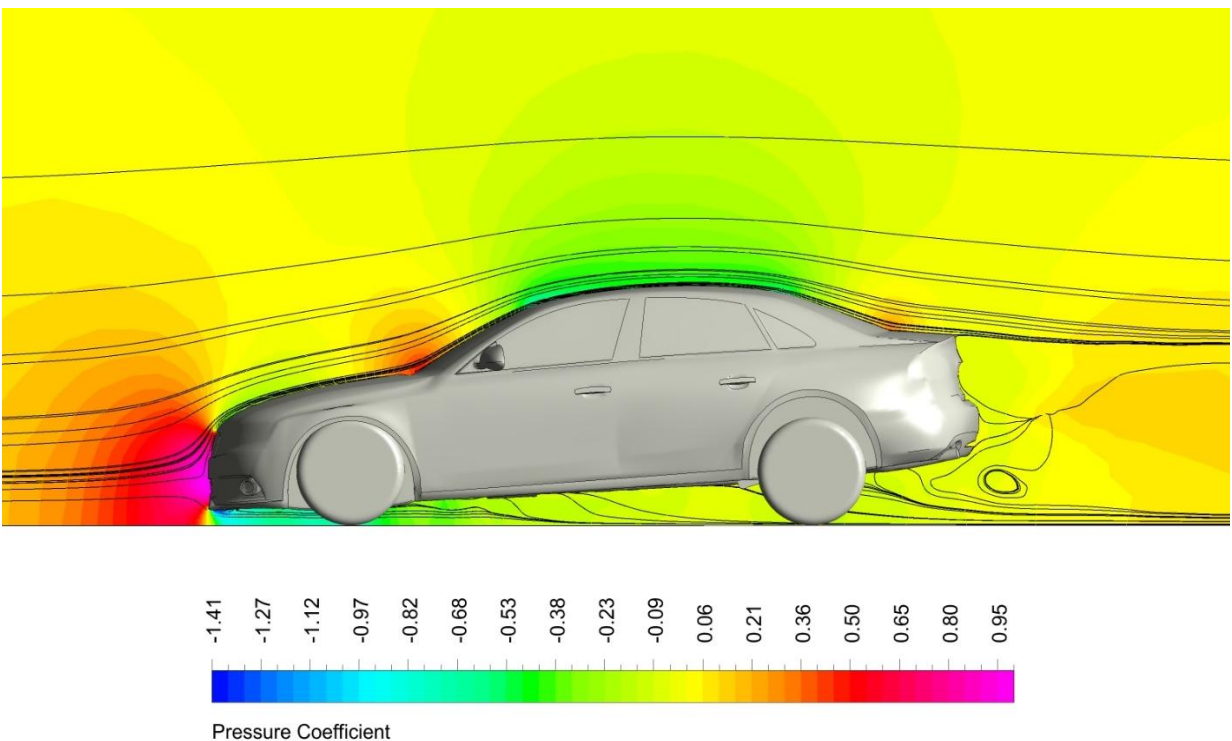

**Figure 15.** Pressure-coefficient ($C_p$) distribution with streamlines on the symmetry plane at optimum ride height h=1.355, optimum rake angle =2.968°, and Re=9.75×10⁶

Since the Reynolds number had only a negligible influence on the location of the optimum solutions, the contour and streamline results are presented only for Re=9.75×10⁶. In the balanced optimization condition with a ride height of 1.355m and a rake angle of 2.968° (Figure 15), the pressure coefficient distribution around the vehicle represents a compromise between drag reduction and downforce generation. In the front stagnation region, a moderate positive pressure is observed, while the underbody maintains an intermediate suction level to ensure stability without causing a significant increase in drag. Over the roof, pressure recovery occurs more smoothly, delaying flow separation and forming a wake of medium size. This case illustrates a balanced aerodynamic condition, providing acceptable aerodynamic efficiency along with sufficient vertical stability for everyday driving scenarios.

## 4.7 Predictive Optimization of Aerodynamic Parameters Across Continuous Reynolds Numbers in the Balanced Condition

In the previous section, optimization was carried out for four discrete Reynolds numbers. After training the machine learning models, the same framework was extended to predict optimal aerodynamic configurations continuously across the full range of Reynolds numbers within the design domain. Using the trained models together with the differential evolution (DE) algorithm, the objective function—defined as a weighted balance between drag and lift coefficients—was evaluated under the balanced condition. To investigate how optimal configurations evolve with flow regime, the Reynolds number range from $4.87 \times 10^6$ to $19.48 \times 10^6$ was divided into ten intervals, and the optimization procedure was applied at each representative point.

It is important to note that this range was explicitly defined and implemented in the code, ensuring that the optimization process systematically explored the selected Reynolds numbers. These limitations were intentionally applied to keep the design domain controlled and to ensure that the simulations were performed under reliable and stable conditions. A key finding is that the optimized conditions remained within narrow geometric intervals, showing only very weak dependence on Reynolds number across the considered range.

Figures 16(a)–16(d) present the resulting variations: subfigures (a) and (b) show the optimal drag and lift coefficients across Reynolds number, while subfigures (c) and (d) depict the corresponding optimal ride height and rake angle.The results indicate that from Re = 4.87×10$^6$ up to Re = 19.48×10$^6$, the drag coefficient remains confined to a narrow interval between 0.2870 and 0.2918, while the lift coefficient gradually decreases from −0.0806 to −0.0888. The ride height varies only slightly, from 1.3554 m down to 1.3403 m, and the rake angle shifts within a limited band of 2.88°–2.97°. The overall objective function remains highly stable, fluctuating between 0.1761 and 0.1776.

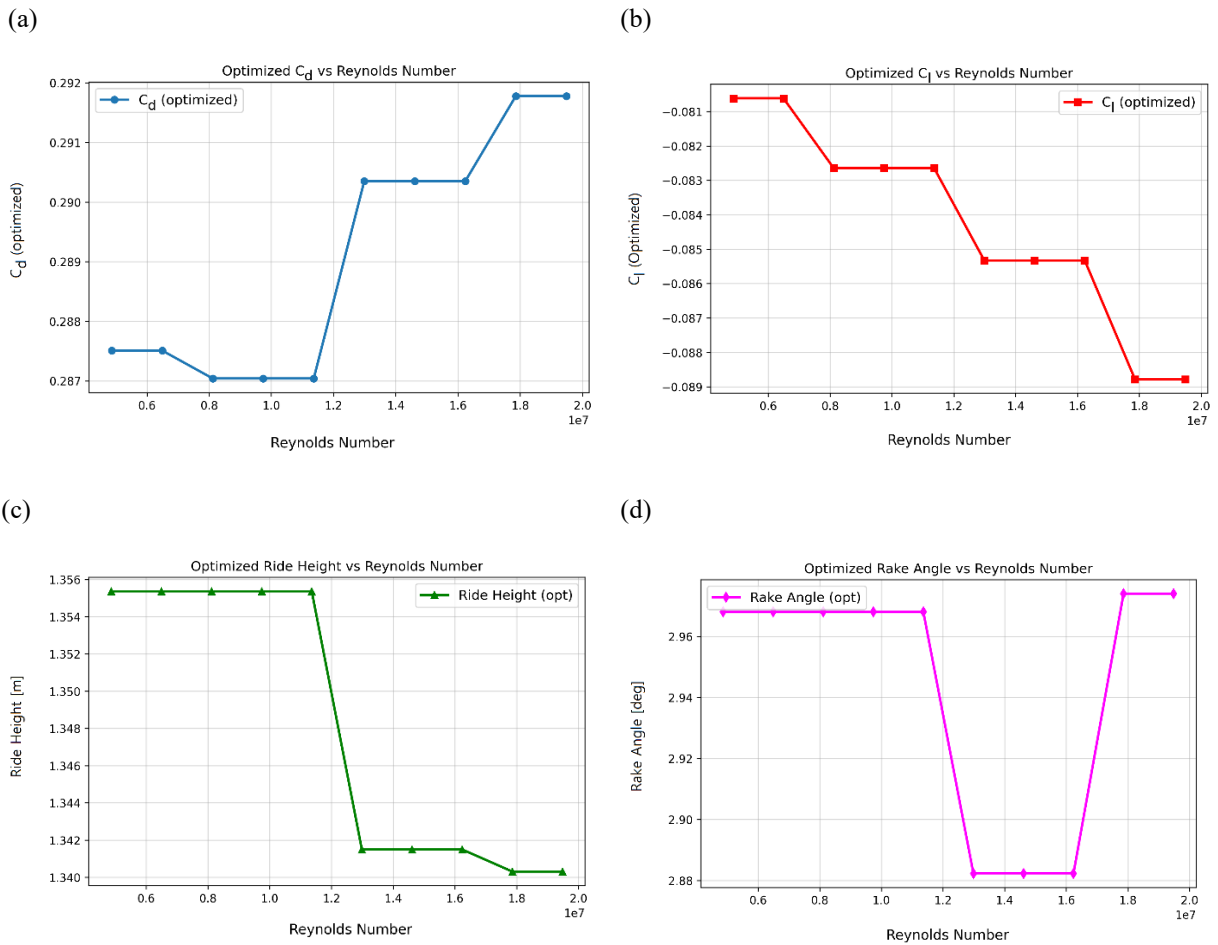

**Figure 16**. Variation of optimal aerodynamic parameters with Reynolds number: (a) drag coefficient versus Reynolds number, (b) lift coefficient versus Reynolds number, (c) optimal ride height versus Reynolds number, (d) optimal rake angle versus Reynolds number in the Balanced Condition

These findings demonstrate that Reynolds number has only a minor effect on the optimal configuration under the balanced condition. The compact range of optimal ride height (≈1.34–1.36 m) and rake angle (≈2.9°) ensures robust aerodynamic performance across different flow regimes, eliminating the need for frequent setup changes. This stability suggests that a narrow optimal design range can consistently maintain favorable aerodynamic performance across a wide range of flow conditions, reducing the necessity for frequent adjustments to vehicle configuration with changing operating speeds and confirming the effectiveness of the optimization approach in capturing a design space that balances drag reduction and downforce over a broad operating domain.

## 4.8 Optimization of Ride Height and Rake Angle for Minimizing Drag Coefficient Across Various Reynolds Numbers

Based on the line plots of $C_d$ and $C_l$ presented in Figures 17(a)–17(h), corresponding to four Reynolds numbers of $4.87 \times 10^6$, $9.75 \times 10^6$, $14.61 \times 10^6$, and $19.48 \times 10^6$ for the Audi A4 model, the optimal ride height and rake angle are clearly identified. Subfigures (a), (c), (e), and (g) show the variation of $C_d$ and $C_l$ with ride height at the optimal rake angle, while subfigures (b), (d), (f), and (h) illustrate the variation with rake angle at the optimal ride height, with optimal points marked on each curve. At Re = $4.87 \times 10^6$, the optimal configuration corresponds to a ride height of 1.3398 m and a rake angle of 0.1588°, resulting in a drag coefficient of 0.28581 and a lift coefficient of 0.01329. The weighted objective function ($0.999 \cdot C_d + 0.001 \cdot C_l$) is 0.28554, indicating a configuration strongly optimized for drag reduction with minimal influence from lift. For Re = $9.75 \times 10^6$, the optimal ride height and rake angle remain nearly unchanged at 1.3408 m and 0.1588°, yielding $C_d$ = 0.28503, $C_l$ = 0.01422, and an objective value of 0.28476, reflecting stable aerodynamic performance across slightly varied flow conditions. At Re = 14.61×10⁶, the ride height increases slightly to 1.3456 m while the rake angle adjusts to 0.2525°, producing $C_d$ = 0.28449, $C_l$ = 0.01317, and an objective value of 0.28422, maintaining near-optimal aerodynamic balance. Finally, at Re = 19.48×10⁶, the optimal configuration is at a ride height of 1.3450 m and a rake angle of 0.3515°, with $C_d$ = 0.28961, $C_l$ = 0.01270, and an objective value of 0.28933, demonstrating that aerodynamic efficiency is sustained even at the highest tested Reynolds number. Overall, the results indicate that the optimal ride height and rake angle remain confined within a narrow range, with drag coefficients consistently around 0.285–0.290 and lift coefficients below 0.015, confirming that a compact set of geometric parameters can ensure near-optimal aerodynamic performance for the Audi A4 model over a wide range of Reynolds numbers.

(a)
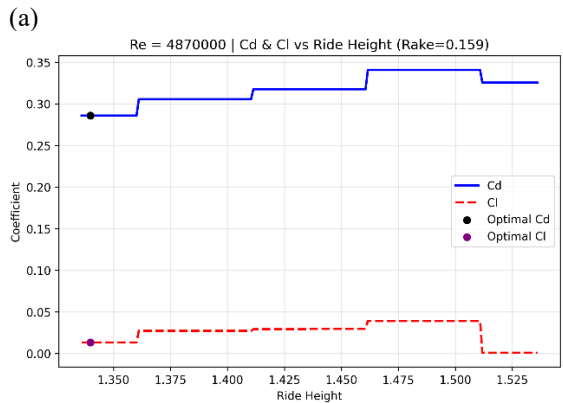

(b)
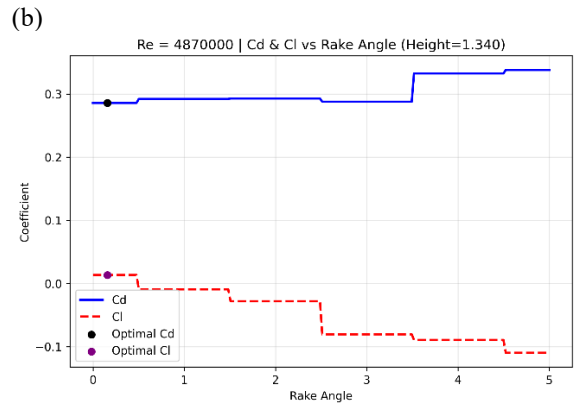

(c)
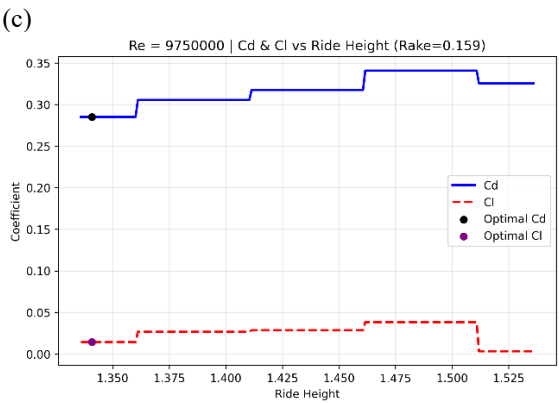

(d)
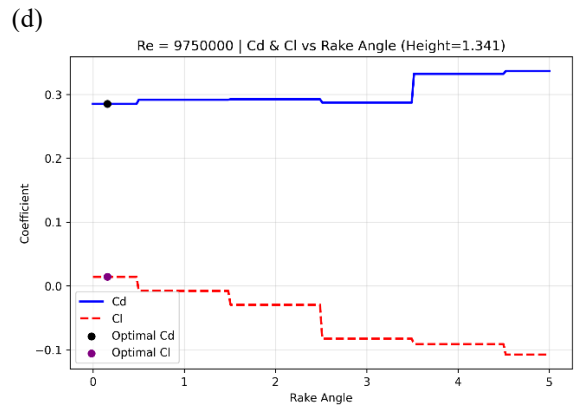

(e)

(f)

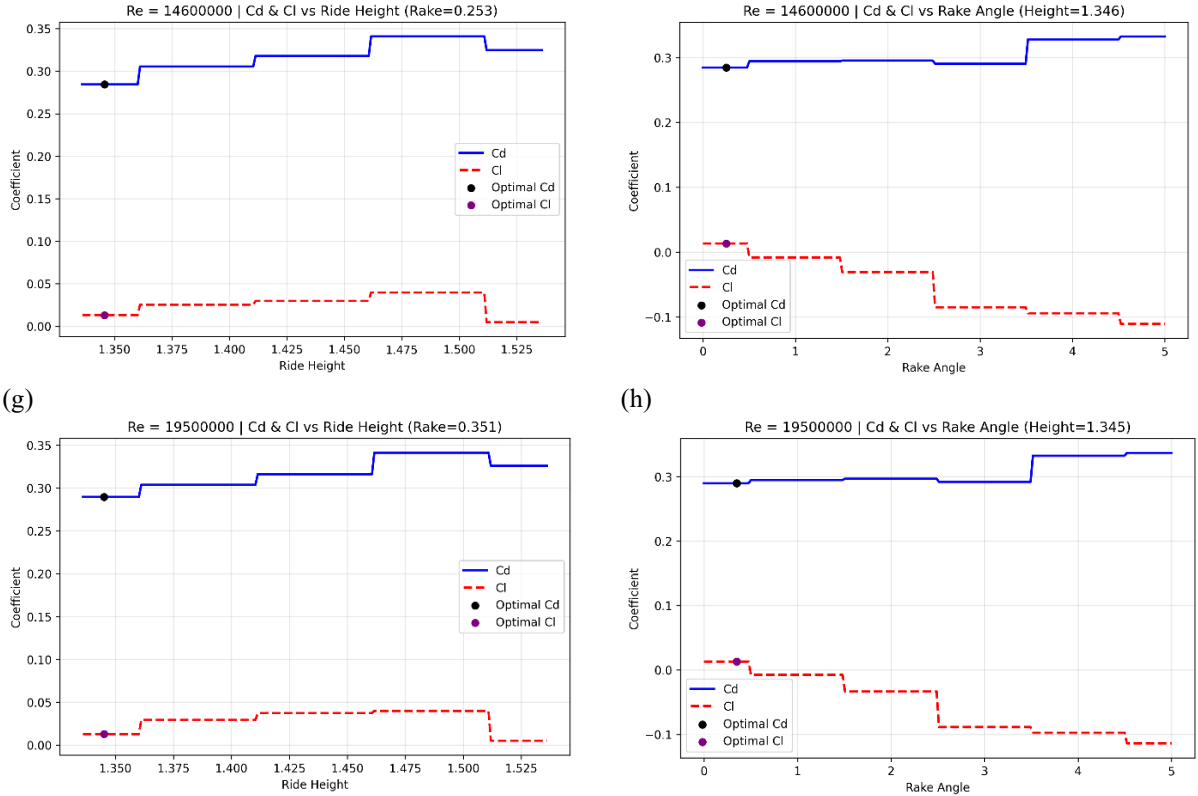

**Figure 17**. Plots of $C_d$ and $C_L$, respectively, based on the combined objective function $(0.999 \cdot C_d + 0.001 \cdot C_l)$ for the Audi A4 at four Reynolds numbers: (a, b) Re = $4.87 \times 10^6$, (c, d) Re = $9.75 \times 10^6$, (e, f) Re = $14.61 \times 10^6$, and (g, h) Re = $19.48 \times 10^6$. For each Reynolds number, the first subfigure [(a), (c), (e), (g)] shows the variation of $C_d$ and $C_l$ with ride height at the optimal rake angle, and the second subfigure [(b), (d), (f), (h)] shows the variation of $C_d$ and $C_l$ with rake angle at the optimal ride height. Optimal points are marked on each curve

For reference, the baseline geometry of the vehicle (ride height = 1.436 m and rake angle = 0°) exhibited drag coefficients between 0.313 and 0.316 and lift coefficients ranging from 0.02 to 0.04 across the four simulated Reynolds numbers. This comparison highlights that the optimization, aimed at minimizing drag, achieved a reduction of approximately 9% in $C_d$, while the lift coefficient increased by roughly 35–70% and became more positive through adjustment of ride height and rake angle.

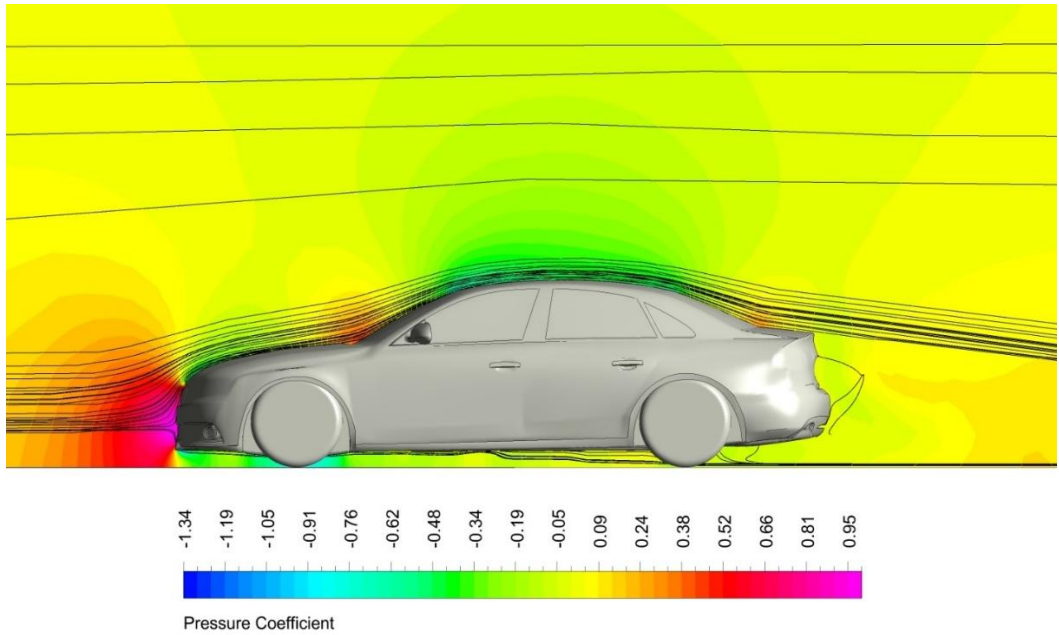

**Figure 18.** Pressure-coefficient ($C_p$) distribution with streamlines on the symmetry plane at optimum ride height h=1.341, optimum rake angle =0.158°, and Re=9.75×$10^6$

Considering the minor influence of Reynolds number on the optimal parameters, the pressure coefficient and streamline patterns are analyzed only at Re=9.75×$10^6$. In the minimum-drag optimization condition with a ride height of 1.341m and a rake angle of 0.158° (Figure 18), the pressure field indicates that the underbody experiences higher pressure compared to the other conditions, which reduces suction and consequently decreases vertical loads. The airflow over the roof exhibits a more uniform pressure gradient, leading to delayed separation and a narrower wake region. As a result, the drag coefficient reaches its minimum value and the aerodynamic efficiency is maximized. This configuration is most favorable for improving fuel economy and long-distance driving, where reducing aerodynamic resistance is prioritized over enhancing downforce.

## 4.9 Optimization of Ride Height and Rake Angle for Maximizing Negative Lift Coefficient Across Various Reynolds Numbers

In this optimization phase, the design target is shifted toward maximum vehicle stability during driving. Here, aerodynamic drag and fuel consumption take a secondary role, while preventing rollover and ensuring safe handling at high speeds are prioritized. The objective function is accordingly tuned with $\alpha = 0.001$ and $\beta = 0.999$, placing almost all weight on maximizing negative lift (downforce), which presses the vehicle firmly toward the road surface [47]. Figures 19(a)–19(h) illustrate the variation of the drag coefficient and lift coefficient for four Reynolds numbers ($4.87 \times 10^6$, $9.75 \times 10^6$, $14.61 \times 10^6$, and $19.48 \times 10^6$) in the Audi A4 model. Subfigures (a), (c), (e), and (g) show $C_d$ and $C_l$ variations with ride height at the optimal rake angle, while subfigures (b), (d), (f), and (h) show the same coefficients as rake angle changes at the optimal ride height. Optimal points are clearly marked on each curve. At Re = $4.87 \times 10^6$, the optimal configuration corresponds to a ride height of 1.3434 m and a rake angle of 4.7852°, giving $C_d = 0.33790$ and $C_l$ = -0.10986. The resulting objective function value of -0.10941 reflects strong downforce generation. For Re = $9.75 \times 10^6$, the optimal ride height increases slightly to 1.3651 m and the rake angle adjusts to 4.6105°, yielding $C_d = 0.33444$, $C_l$ = -0.10842, and an objective of -0.10798, indicating stability is maintained with minimal geometric change. At Re = $14.61 \times 10^6$, the ride height slightly decreases to 1.3580 m with a rake angle of 4.7645°, giving $C_d = 0.33230$, $C_l$ = -0.11098, and an objective of -0.11053. Finally, at Re = $19.48 \times 10^6$, the configuration remains at 1.3580 m and 4.7645°, with $C_d = 0.33655$, $C_l$ = -0.11401, and an objective of -0.11356.

(a)                                             (b)

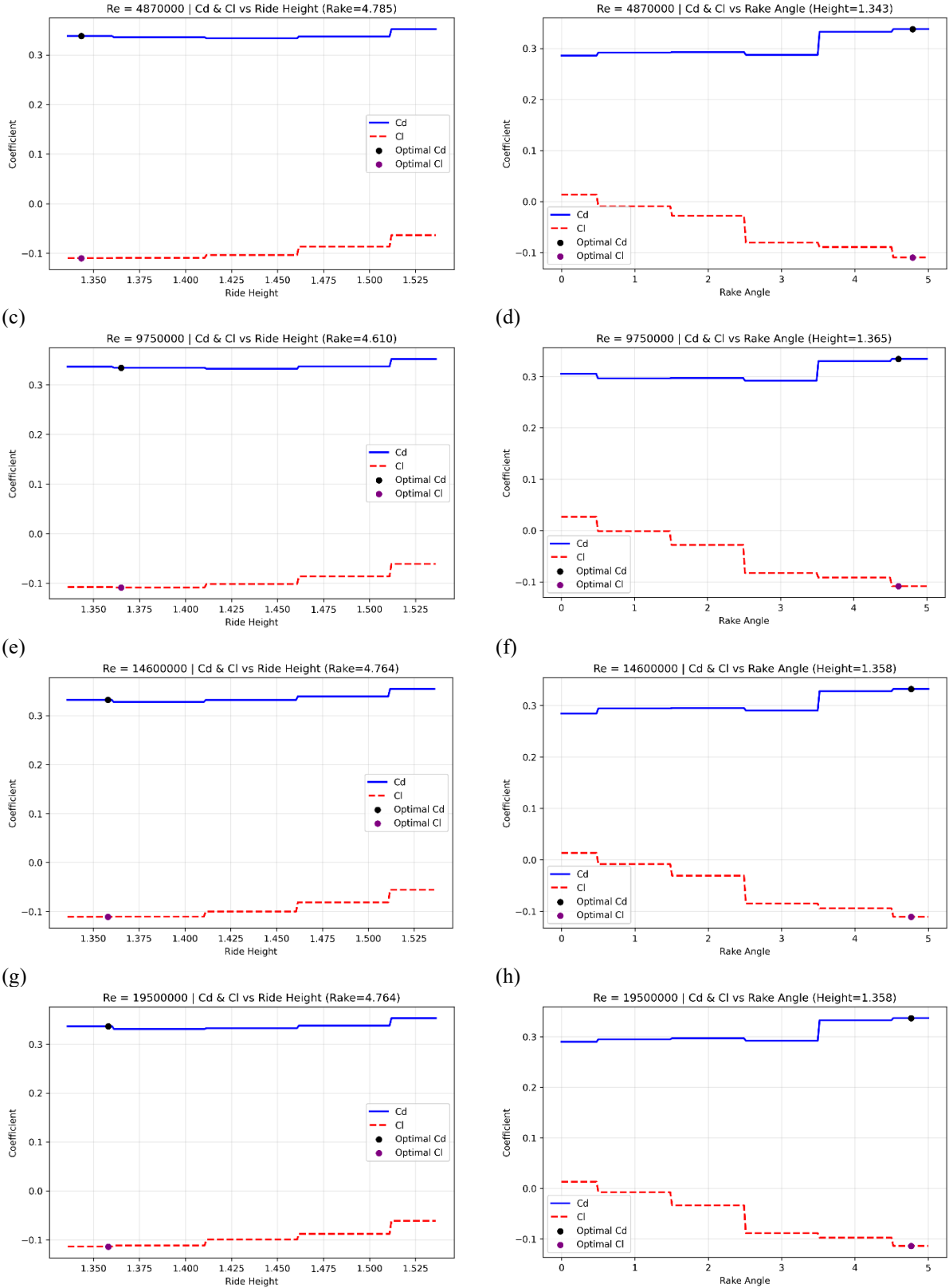

**Figure 19**. Plots of $C_d$ and $C_l$, respectively, based on the combined objective function $(0.001 \cdot C_d + 0.999 \cdot C_l)$ for the Audi A4 at four Reynolds numbers: (a, b) Re = $4.87 \times 10^6$, (c, d) Re = $9.75 \times 10^6$, (e, f) Re = $14.61 \times 10^6$, and (g, h) Re = $19.48 \times 10^6$. For each Reynolds number, the first subfigure [(a), (c), (e), (g)] shows the variation of $C_d$ and $C_l$ with ride height at the Comparing these optimized conditions with the baseline setup (ride height = 1.436 m and rake angle = 0°), which exhibited $C_d$ between 0.313–0.316 and $C_l$ between 0.02–0.04, reveals significant aerodynamic changes. The drag coefficient increased by approximately 6–9%, while the lift coefficient became more negative, increasing downforce by roughly 200–650% depending on Reynolds number. This confirms that prioritizing stability through adjustment of ride height and rake angle can substantially enhance vehicle downforce while increasing aerodynamic drag, providing safer handling at high speeds without excessive compromise on overall aerodynamic performance.

Overall, these results show that both optimal ride height and rake angle remain in a narrow range across different flow conditions, with $C_l$ consistently near -0.10. This confirms that a compact set of geometric parameters can deliver almost ideal downforce and stability for the Audi A4, an outcome especially valuable in high-speed racing or other critical driving scenarios [60].

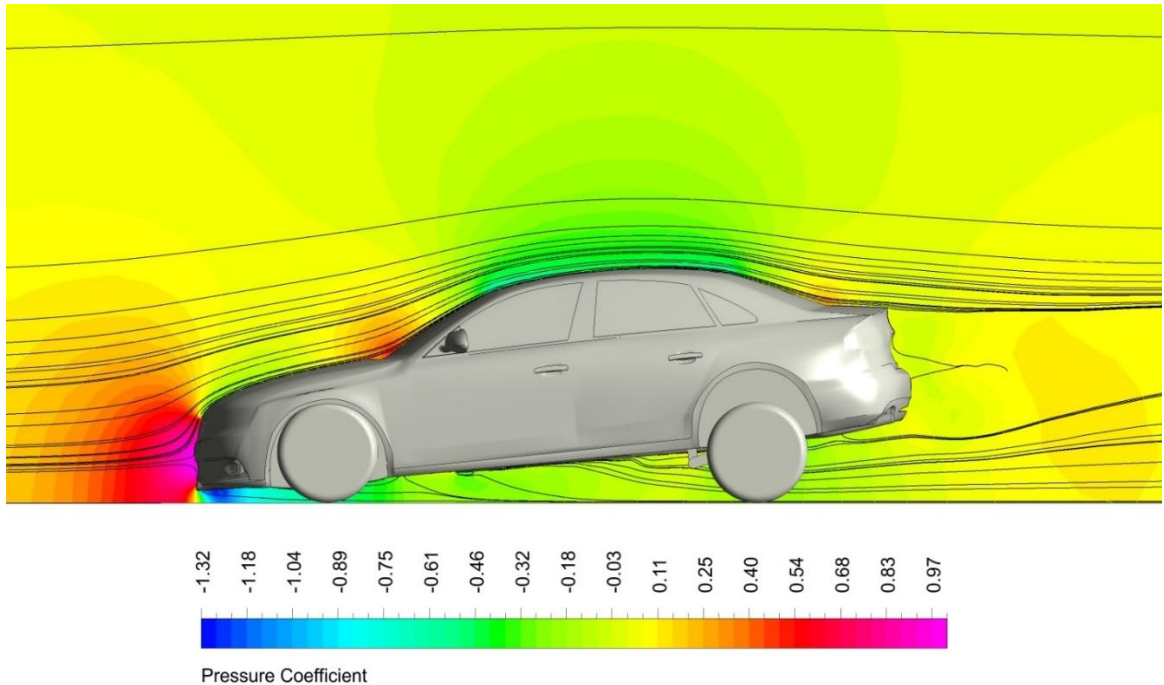

**Figure 20**. Pressure-coefficient ($C_P$) distribution with streamlines on the symmetry plane at optimum ride height h=1.365, optimum rake angle =4.610°, and Re=9.75×10^6

As the variation of Reynolds number showed only a limited effect on the optimal conditions, the pressure distribution and streamlines are illustrated solely for Re=9.75×10^6. In the maximum negative-lift optimization condition with a ride height of 1.365m and a rake angle of 4.610° (Figure 20), the pressure beneath the vehicle is significantly reduced, producing strong suction and consequently a considerable increase in downforce. In the front stagnation region, the pressure is slightly higher, reinforcing the pressure difference between the upper and lower surfaces of the vehicle. However, this configuration also results in a larger wake region behind the vehicle, characterized by stronger flow separation and weaker pressure recovery, which leads to increased drag. This case enhances stability and road holding at high speeds, but it does so at the expense of aerodynamic efficienc

## 4.10 Validation of Aerodynamic Optimization

According to the results obtained in this study, since no significant differences were observed in the optimal values across different Reynolds numbers, the flow analysis was performed at a single Reynolds number, representing the other conditions. This stability in the optimal ride height and rake angle can be attributed to the aerodynamic flow characteristics within the investigated range. Although an increase in Reynolds number slightly reduces boundary layer thickness and causes minor shifts in the flow separation point, these effects are too small to significantly alter the overall flow pattern around the vehicle. Consequently, the geometric configuration that produces optimal aerodynamic performance remains stable for most Reynolds numbers considered, with only minor adjustments required to maintain optimal conditions

To evaluate the aerodynamic optimization performance, Table 3 presents the predicted results of the Gradient Boosting algorithm at a Reynolds number of Re=$9.75 \times 10^6$. In this table, three optimized conditions (balanced, minimum drag, and maximum negative lift) are compared with the baseline geometry of the vehicle without an adaptive control system. As shown, in the balanced optimization condition with a ride height of 1.355m and a rake angle of 2.968°, the drag coefficient is reduced by 8.306% relative to the baseline, while a significant increase in negative lift (approximately 386%) is achieved. In the minimum-drag optimization condition with a ride height of 1.341m and a rake angle of 0.158°, the highest drag reduction is obtained (about 8.945%), while the change in negative lift is relatively smaller (around 50%). Finally, in the maximum negative-lift optimization condition with a ride height of 1.365m and a rake angle of 4.610°, the negative lift is greatly enhanced (approximately 476% compared to the baseline), but at the cost of a 6.709% increase in drag.

**Table 3**. Predicted aerodynamic coefficients and their errors relative to the baseline geometry under different optimized conditions for Re=9.75×10⁶

| Conditions | Ride Height (m) | Rake Angle (deg) | $C_d$_Predicted | $C_l$_Predicted | $\frac{\Delta C_d}{C_d} \times 100$ | $\frac{\Delta C_l}{C_l} \times 100$ |
|---|---|---|---|---|---|---|
| Baseline Geometry | 1.436 | 0 | 0.313 | +0.0288 | 0 | 0 |
| Optimized balanced | 1.355 | 2.968 | 0.287 | -0.0826 | -8.306 | -386.805 |
| Minimum Drag | 1.341 | 0.158 | 0.285 | +0.0142 | -8.945 | -50.694 |
| Maximum Downforce | 1.365 | 4.610 | 0.334 | -0.1084 | +6.709 | -476.389 |

These findings demonstrate that, depending on the design objective, an appropriate balance between reducing fuel consumption (through drag reduction) and improving high-speed stability (through increased negative lift) can be established by adjusting ride height and rake angle. Therefore, the results in Table 3 confirm that the Gradient Boosting algorithm effectively identified and predicted three distinct aerodynamic optimization conditions in comparison with the baseline geometry.

### 4.11 Evaluation of Machine Learning Predictions with CFD Simulations

To further evaluate the aerodynamic performance at the three optimized conditions (Balanced, Minimum Drag, and Maximum Downforce), the accuracy of the machine learning predictions was assessed against CFD simulations. The results indicate that the relative error in the drag coefficient lies within approximately 0.9–2.5%, whereas the relative error in the lift coefficient is around 9–10%. This trend is expected, since $C_d$ represents an integrated and relatively stable quantity that can be more reliably predicted by surrogate models, while $C_l$ is highly sensitive to local pressure variations and separation details, which amplifies the relative error [1, 2, 61, 62]. It should also be noted that all CFD analyses were carried out at a Reynolds number of 9.75×10⁶. Since the influence of Reynolds number on the location of the optimized conditions was found to be negligible, all results are consistently reported at this single Reynolds condition.

Subsequently, each optimized geometry was simulated in CFD, and the resulting aerodynamic coefficients are reported in Table 4. The Balanced condition yields $C_d = 0.293$ and $C_l = -0.062$. The Minimum Drag condition reduces drag to $C_d = 0.278$ but produces slightly positive lift ($C_l = 0.013$). In contrast, the Maximum Downforce condition increases drag to $C_d = 0.331$ while achieving the strongest negative lift with $C_l = -0.12$. It should be noted that the higher relative error in $C_l$, although expected, indicates that optimization based on the surrogate model may lead to solutions close to, but not exactly, the true optimum—particularly for objectives focused on downforce. This highlights the importance of final validation using CFD for each candidate design identified by the machine learning framework.

Table 4. Comparison of predicted and simulated aerodynamic coefficients with percentage errors for optimized conditions.

| Conditions | $C_d$_Simulated | $C_l$_Simulated | $\left\| \dfrac{\Delta C_d}{C_d\_\text{Simulated}} \times 100 \right\|$ | $\left\| \dfrac{\Delta C_l}{C_l\_\text{Simulated}} \times 100 \right\|$ |
|---|---|---|---|---|
| Optimized balanced | 0.293 | -0.062 | 2.04 | 9.10 |
| Minimum Drag | 0.278 | 0.013 | 2.518 | 9.23 |
| Maximum Downforce | 0.331 | -0.12 | 0.906 | 9.67 |

Flow-field inspection clarifies the physical mechanisms behind these outcomes. In the Maximum Downforce condition, the centerline $C_p$ distribution along the underbody (Figure 21(b)) shows a pronounced pressure drop in the front and mid-floor regions, indicative of stronger suction beneath the vehicle. This intensified pressure differential between the lower and upper surfaces (Figure 21(a)) directly generates larger negative lift but also enlarges the wake and increases drag [1,2]. Conversely, the Minimum Drag condition exhibits higher static pressure under the body and a smoother pressure recovery over the roof and rear surfaces, which suppress wake size and reduce overall drag. However, the reduced underbody suction eliminates downforce and even results in slightly positive lift [3]. The Balanced condition demonstrates intermediate characteristics: underbody suction is stronger than in the Minimum Drag condition but less extreme than in the Maximum Downforce condition, while the pressure gradient over the upper body remains smoother than in the Maximum Downforce condition. This compromise explains why the Balanced design achieves a moderate drag reduction while maintaining a small amount of negative lift.

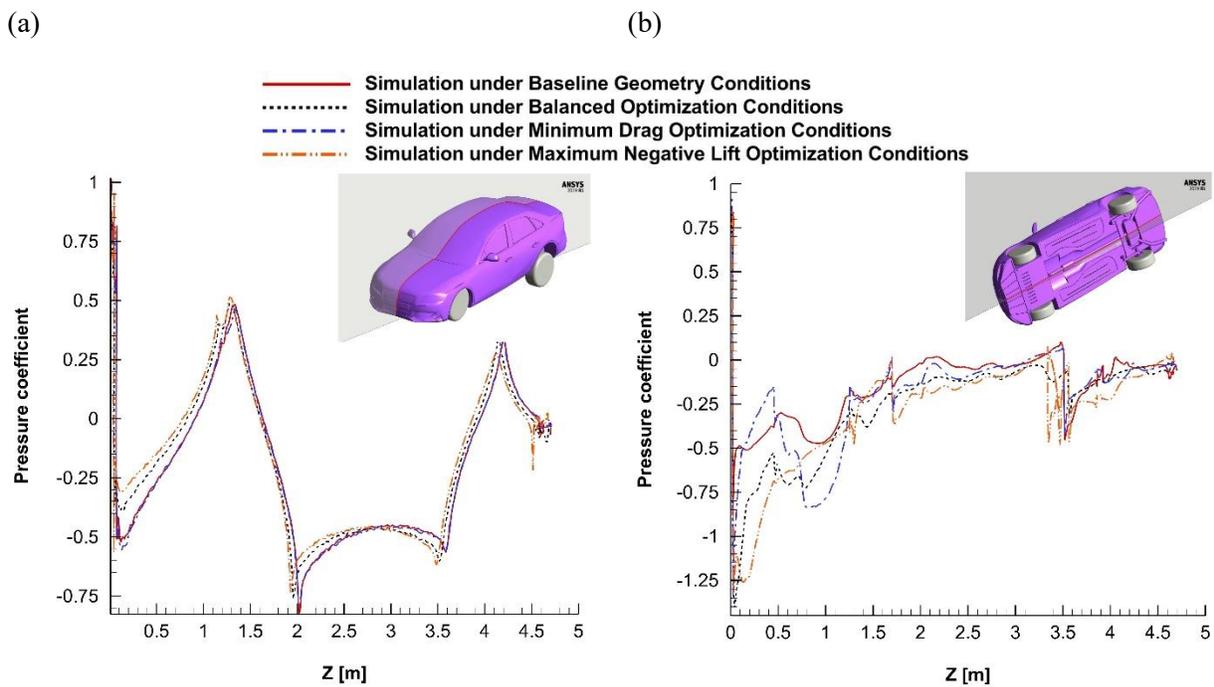

**Figure 21**. The distribution of pressure coefficient (a) on the body and (b) under the body along the symmetry plane is simulated for three optimized conditions and the baseline geometry condition in a flow with Reynolds number $9.75 \times 10^6$

Overall, the CFD analysis confirms that the aerodynamic trade-offs among the three optimized geometries can be directly traced to distinct pressure distributions on the upper and lower body surfaces. Stronger underbody suction enhances downforce at the cost of increased drag, whereas smoother upper-body pressure recovery reduces drag but limits downforce generation. These flow-structure mechanisms are consistent with prior investigations of ride height and underbody aerodynamics in passenger vehicles [1, 2, 61, 62], indicating that the observed tendencies are physically robust rather than coincidental.

## 4.12 Evaluation of the Impact of Aerodynamic Optimization on Fuel Consumption and Stability

The aerodynamic optimization of vehicle geometry significantly influences both stability and fuel efficiency [22]. To quantify these effects, a force and energy balance analysis was conducted for three optimized conditions (balanced, minimum drag, and maximum downforce) in comparison with the baseline geometry at a Reynolds number of $9.75 \times 10^6$. The vehicle mass was assumed constant (m=1600 kg) [63], and wheel friction was represented only through the rolling resistance coefficient ($C_r$=0.01) [64]. Aerodynamic forces were determined based on the drag and lift coefficients obtained from CFD simulations [47] (Eq. 10,11). The required engine force, $F_{engine}$, was calculated as the sum of aerodynamic drag and rolling resistance (Eq. 12), while the total energy consumption, $E_{total}$, was obtained by multiplying $F_{engine}$ by a displacement of 100 km (Eq. 13) [65]. Effective fuel energy, $E_{fuel}$, was then derived from $E_{total}/\eta_{engine}$ with an assumed engine efficiency of 0.3 (Eq. 14) [63], and the fuel volume consumption, $V_{fuel}$, was computed by dividing $E_{fuel}$ by the specific energy of gasoline (33 MJ/L) (Eq. 15) [66].

$$D = C_d \times (\frac{1}{2} \rho U_\infty^2 A_{front}) \tag{10}$$

$$L = C_l \times (\frac{1}{2} \rho U_\infty^2 A_{top}) \tag{11}$$

$$F_{engine} = C_r(mg - L) + D \tag{12}$$

$$E_{total} = F_{engine}.\Delta x \tag{13}$$

$$E_{fuel} = \frac{F_{engine}.\Delta x}{\eta_{engine}} \tag{14}$$

$$V_{fuel} = \frac{E_{fuel}}{E_{per\ liter}} \tag{15}$$

According to Figure 22, the baseline geometry produced a small positive lift of 127.5 N, indicating a tendency toward upward aerodynamic force. Under the balanced optimization, lift shifted substantially to −273.9 N, which corresponds to a significant enhancement in downforce and consequently improved stability. In the minimum drag condition, lift was moderately reduced to 57.6 N, whereas the maximum downforce condition generated the highest stability margin with −528.6 N of lift, more than four times greater downforce compared to the baseline. These results demonstrate that aerodynamic stability improvement is most pronounced in the maximum downforce condition, followed by the balanced condition [67].

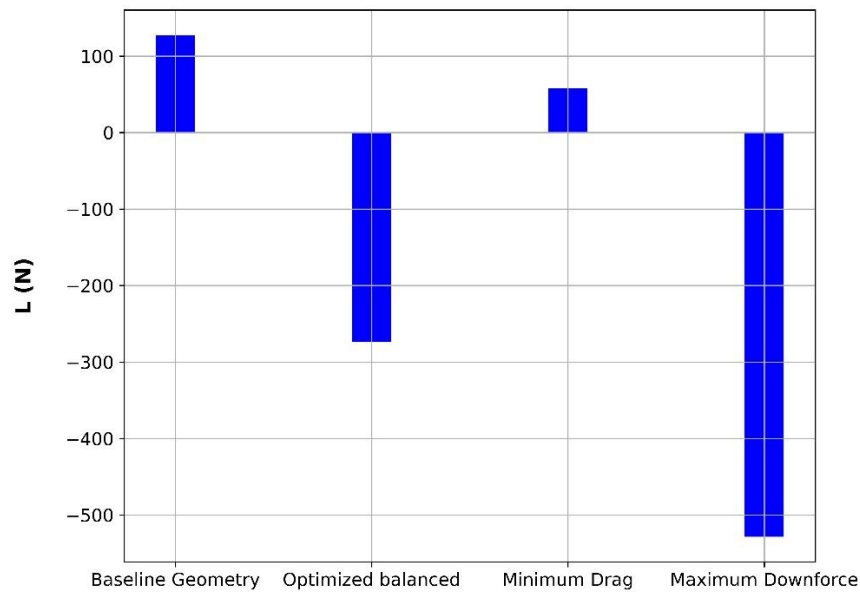

In terms of propulsive demand (Figure 23), the baseline configuration required 532.5 N of engine force. The minimum drag condition reduced this requirement to 483.9 N, corresponding to a 9.1% improvement in efficiency. Conversely, the balanced condition slightly increased the demand to 533.9 N, which is practically unchanged compared to the baseline. The maximum downforce condition, however, imposed a substantial penalty, raising the required engine force to 608.7 N (14.3% higher than baseline) as a consequence of increased aerodynamic resistance.

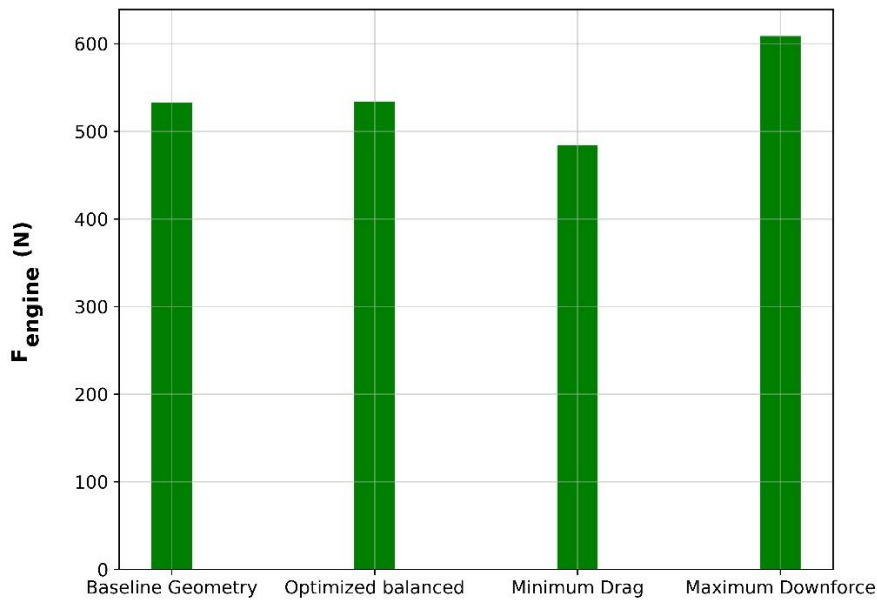

**Figure 23**. Required engine force ($F_{engine}$) for baseline and optimized conditions

Fuel energy consumption followed the same pattern (Figure 24). The baseline geometry required 177.5 MJ to cover 100 km, whereas the minimum drag condition reduced this to 161.3 MJ (9.1% reduction). The balanced condition showed a negligible increase to 178.0 MJ, while the maximum downforce condition reached 202.9 MJ, which reflects a 14.3% higher demand compared to the baseline.

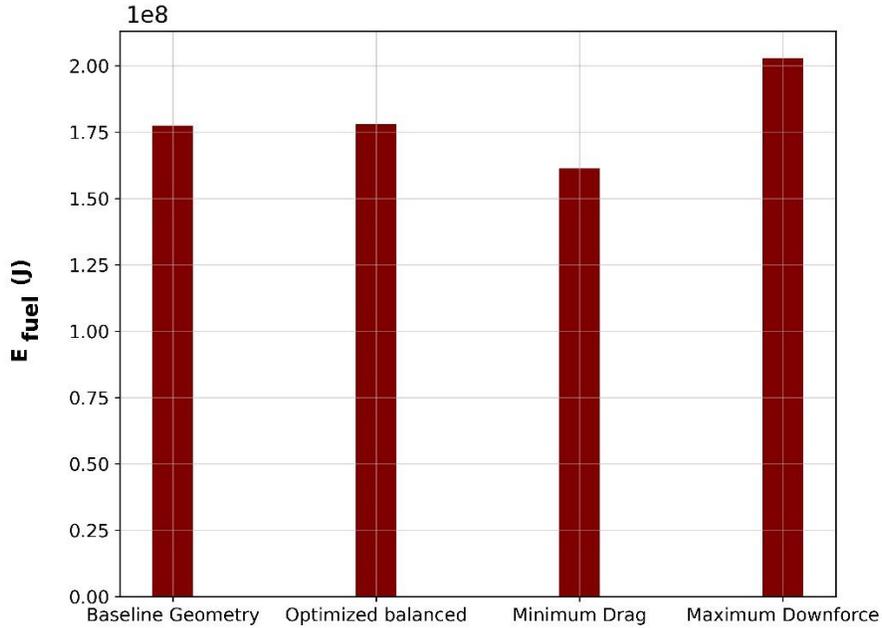

**Figure 24**. Fuel energy consumption ($E_{fuel}$) over 100 km for different aerodynamic conditions

The translation of these energy values into fuel volume shows that (refer to Figure 25) the baseline condition required 5.38 L/100 km. This slightly increased to 5.39 L/100 km in the balanced condition (+0.26%). In the minimum drag condition, fuel demand decreased to 4.89 L/100 km (−9.13%), whereas the maximum downforce condition increased fuel consumption to 6.15 L/100 km (+14.31%) [47,67].

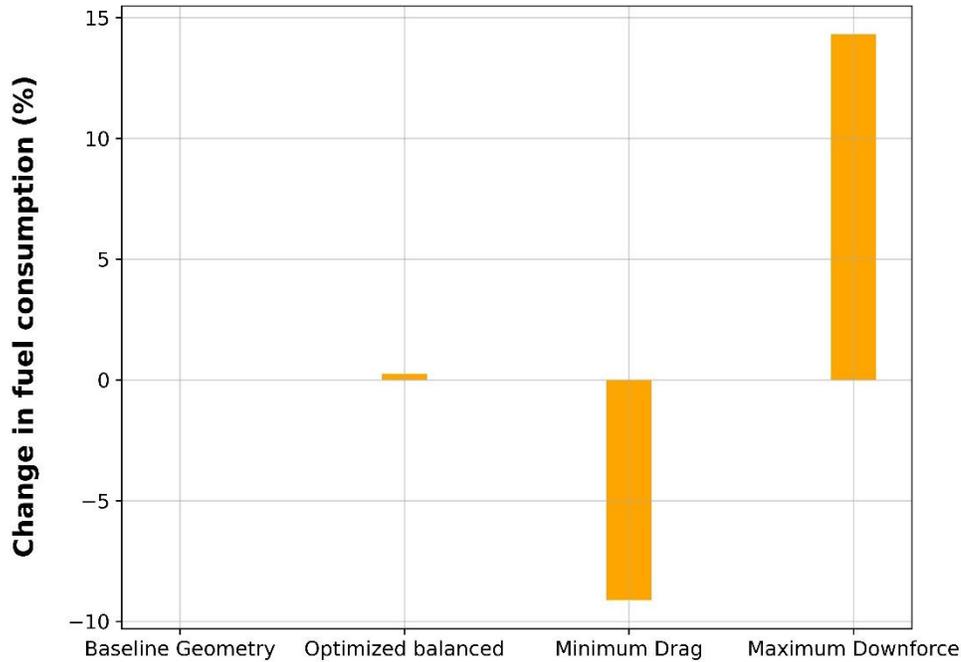

**Figure 25**. Percentage change in fuel consumption ($\Delta V/V$ %) compared to baseline geometry

In conclusion, aerodynamic optimization leads to markedly different outcomes depending on the design target. The minimum drag condition maximizes fuel economy by reducing drag forces, whereas the balanced optimization achieves a compromise between enhanced downforce and unchanged energy efficiency. On the other hand, the maximum downforce condition provides the highest stability but at the expense of a substantial increase in energy consumption. Accordingly, the choice of an optimal aerodynamic configuration should be guided by the intended operational priority, whether fuel economy or vehicle stability [22, 47].

## 5. Conclusion

This study demonstrated that ride height and rake angle are decisive factors in shaping the aerodynamic performance of passenger vehicles. Because Reynolds number showed only a minor influence on the optimal configuration, detailed quantitative results were reported for Re = $9.75\times10^6$, while other conditions followed similar trends. Using CFD simulations combined with Gradient Boosting surrogate models and Differential Evolution optimization, three aerodynamic conditions were identified. In the balanced condition, a ride height of ≈1.35 m and a rake angle of ≈2.9° yielded a ~7–8% drag reduction while improving downforce by ~386%. In the minimum-drag condition, drag was reduced by ~9% at the cost of slightly positive lift. In the maximum downforce condition, negative lift increased by ~476%, though drag rose by ~7%. These results confirm the trade-off between fuel efficiency and high-speed stability.

A key outcome is that optimal configurations remained confined to narrow geometric ranges (ride height ≈1.34–1.36 m, rake ≈0.16–4.8°) with very weak dependence on Reynolds number, allowing a single set of adjustments to deliver robust performance across varied speeds. This finding supports the feasibility of adaptive suspension systems that dynamically control vehicle posture to maintain aerodynamic efficiency in real-world conditions.

Furthermore, machine-learning predictions were externally validated by CFD simulations, achieving less than 3% error for drag coefficient, thereby confirming the reliability of the surrogate approach. Beyond aerodynamic coefficients, energy and force analyses revealed that optimization can reduce fuel consumption by up to 9.1% in the minimum-drag condition or enhance downforce by more than 500% in the maximum stability condition. Thus, the presented framework not only provides a computationally efficient path for aerodynamic design but also offers practical solutions to balance efficiency, handling, and safety under diverse driving conditions.

Despite the valuable outcomes of this study, several limitations should be addressed in future research. First, the effect of crosswind was not considered, although it can significantly influence vehicle aerodynamic performance in real-world conditions. Second, despite the complexity of the model, certain geometric simplifications were made, such as omitting detailed airflow modeling within the engine compartment and the complete underbody. Third, the simulations were conducted under steady-state conditions, whereas transient phenomena such as vortex shedding could be explored using more advanced and computationally expensive methods like DES or LES for deeper flow insights. Finally, from a practical perspective, implementing the results through adaptive suspension systems requires addressing challenges such as the response speed needed for adjusting ride height and rake angle, as well as assessing the energy cost of such systems. This analysis can bridge the gap between numerical optimization and real-world engineering applications.

**Data Availability**

The datasets used and/or analyzed during the current study are available from the corresponding author upon reasonable request.